\documentclass[pra,twocolumn,showpacs]{revtex4-1}
\usepackage{amsmath,amscd,amssymb,color}
\usepackage{graphicx,amsfonts,dsfont}
\usepackage{epstopdf}
\usepackage{hyperref}
\usepackage{enumerate,bbold}
\usepackage{array}
\newcommand{\ie}{\textit{i.e.}}
\begin{document}
\title{Simulating open quantum dynamics on an NMR quantum
processor using the Sz.-Nagy dilation algorithm}
\author{Akshay Gaikwad}
\email{ph16010@iisermohali.ac.in}
\affiliation{Department of Physical Sciences, Indian
Institute of Science Education \& 
Research Mohali, Sector 81 SAS Nagar, 
Manauli PO 140306 Punjab India.}
\author{Arvind}
\email{arvind@iisermohali.ac.in}
\affiliation{Department of Physical Sciences, Indian
Institute of Science Education \& 
Research Mohali, Sector 81 SAS Nagar, 
Manauli PO 140306 Punjab India.}
\author{Kavita Dorai}
\email{kavita@iisermohali.ac.in}
\affiliation{Department of Physical Sciences, Indian
Institute of Science Education \& 
Research Mohali, Sector 81 SAS Nagar, 
Manauli PO 140306 Punjab India.}
\begin{abstract}
We experimentally implement the Sz.-Nagy dilation algorithm to
simulate open quantum dynamics on an 
nuclear magnetic resonance (NMR) quantum processor.  The Sz.-Nagy
algorithm enables the simulation of the dynamics of arbitrary-dimensional open
quantum systems, using only a single ancilla qubit.  We experimentally simulate
the action of two non-unitary processes, namely, a phase damping channel acting
independently on two qubits  and  a magnetic field gradient pulse (MFGP) acting
on an ensemble of two coupled nuclear spin-1/2 particles.  To evaluate the
quality of the experimentally simulated quantum process, we perform convex
optimization-based full quantum process tomography to reconstruct the quantum
process from the experimental data and compare it with the target quantum
process to be simulated.  
\end{abstract} 
\maketitle 
\section{Introduction}
\label{sec1}
In 1982, Richard Feynman proposed the idea of simulating quantum systems using a
universal quantum computer~\cite{feynman-ijtp-1982}, which received a lot of
attention from the scientific community~\cite{lloyd-science-1996,
kassal-arpc-2011,franco-rmp-2014}.  Over the following decades, this led to
efforts to build quantum computers which could solve computational problems
exponentially faster, as compared to their classical
counterparts~\cite{nielsen-book-10}.  The main building block of a quantum
computer is the underlying physical system and its time evolution under a given
Hamiltonian~\cite{divi-fdp-2000}, while the main obstacle in building such a
quantum computer is its unwanted and inevitable interaction with its
environment, generally referred to as decoherence~\cite{harper-np-2020}.  This
led to studies of open quantum dynamics, whereby the time evolution of a quantum
system was studied using different approaches~\cite{Breuer2007,Rotter-rpp-2015}.
         
The physical implementation of quantum algorithms mostly relies on unitary
quantum gates. However, in real situations, the physical system under
consideration is continuously interacting with its environment, causing its time
evolution to be non-unitary.  In some cases, the noise from open dynamics can
contribute significantly to errors in the computational output, leading to lower
experimental fidelity and a reduction in the quality of the quantum
device~\cite{zuniga-pra-2012}.  A duality quantum algorithm for simulating
Hamiltonian evolution of an open quantum system was proposed where the time
evolution is realized using Kraus operators~\cite{Wei-sr-2016,Zheng-sr-2021}.  A
quantum algorithm was proposed to simulate a general finite-dimensional Lindblad
master equations without needing to engineer system-environment
interactions~\cite{Candia-sr-2015}.  A method for efficient quantum simulation
of open quantum dynamics for various Hamiltonians and spectral densities was
recently proposed~\cite{Zhang-fp-2021}.

Several techniques have been proposed to simulate specific types of quantum
channels and have been experimentally realized using different physical
platforms.  A control technique to drive an open quantum system from the
Markovian to the non-Markovian regime was demonstrated on an optical
setup~\cite{Liu-np-2011}.  A model was designed that precisely controls the
strength of non-Markovian effects by changing the degree of correlation and
qubit-environment interaction time on an NMR system~\cite{Bernardes-sr-2016}.
Non-positive dynamical maps the decoherence dynamics of a qubit were
experimentally demonstrated using photons~\cite{Liu-natcom-2018}.  A technique
to simulate Markovian and non-Markovian dynamics was proposed on a cavity-QED
setup~\cite{patsch-prr-2020}.  
Multiqubit open dynamics was
simulated on an IBM quantum processor for several quantum processes including
unital and non-unital dynamics as well as Markovian and non-Markovian
evolution~\cite{garcia-npj-2020}.
A dilation
procedure was employed to simulate non-hermitian Hamiltonian dynamics using
ancilla qubits~\cite{dogra-commphy-2021}.

Recently, promising quantum algorithms to simulate arbitrary non-unitary
evolutions on quantum devices have been reported, which are primarily based on
the dilation technique namely, the  Stinespring dilation
algorithm~\cite{shirokov-jmp-2020} and Sz.-Nagy's dilation
algorithm~\cite{prineha-prr-2021}.  The basic tenet of these algorithms is to
construct a unitary operation in a higher-dimensional Hilbert space, which
simulates the desired non-unitary evolution in a lower-dimensional Hilbert
space.  The Stinespring dilation algorithm requires a larger Hilbert space
dimension, which makes it computationally and experimentally expensive, as
compared to the Sz.-Nagy algorithm.  The Sz.-Nagy algorithm has been used to
experimentally simulate the single-qubit amplitude damping channel on the IBM
quantum processor~\cite{Hu-sr-2020}.

In this work, we experimentally implemented the Sz.-Nagy quantum algorithm to
simulate open quantum dynamics on an ensemble NMR quantum information processor.
In order to simulate the given quantum dynamics of an open quantum system, the
Sz.-Nagy algorithm requires prior knowledge of the corresponding complete set of
Kraus operators.  However, in a realistic scenario, the Kraus operators might
not directly available. In such cases, one has to first compute the complete set
of Kraus operators before proceeding with the implementation of the Sz.-Nagy
algorithm.  We used process tomography to first compute the process matrix
which characterizes the given quantum 
process~\cite{gaikwad-pra-2018}. Using unitary diagonalization, we
then compute the complete set of Kraus operators corresponding to a general
quantum channel, using Lindblad generators.  To demonstrate the efficacy of the
Sz.-Nagy algorithm, we experimentally simulated two non-unitary quantum
processes acting on a two-qubit system: a phase damping channel acting
independently on the two qubits where the Kraus operators are already known, and
a magnetic field gradient pulse (MFGP), where the Kraus operators are not
directly available and need to be computed.  Further, to validate the quality of
the experimentally simulated quantum channel, we perform convex
optimization-based full quantum process
tomography~\cite{gaikwad-ijqi-2020,gaikwad-arxiv-2021}.  

This paper is organized as follows: The details of the Sz.-Nagy dilation
algorithm are given in Section~\ref{sec2}.  
The details of using the Sz.-Nagy algorithm to simulate two-qubit
non-unitary quantum processes are given in Section~\ref{sec3}, with
the experimental parameters detailed in Section~\ref{expt}.
The experimental implementations of
the Sz.-Nagy algorithm to simulate an independent phase damping channel 
and to simulate an
MFGP acting on two NMR qubits are described in Section~\ref{pd} 
and \ref{mfgp}, respectively. 
Section~\ref{sec5} contains a few concluding remarks.

\begin{figure}[h] 
\includegraphics[angle=0,scale=1.1]{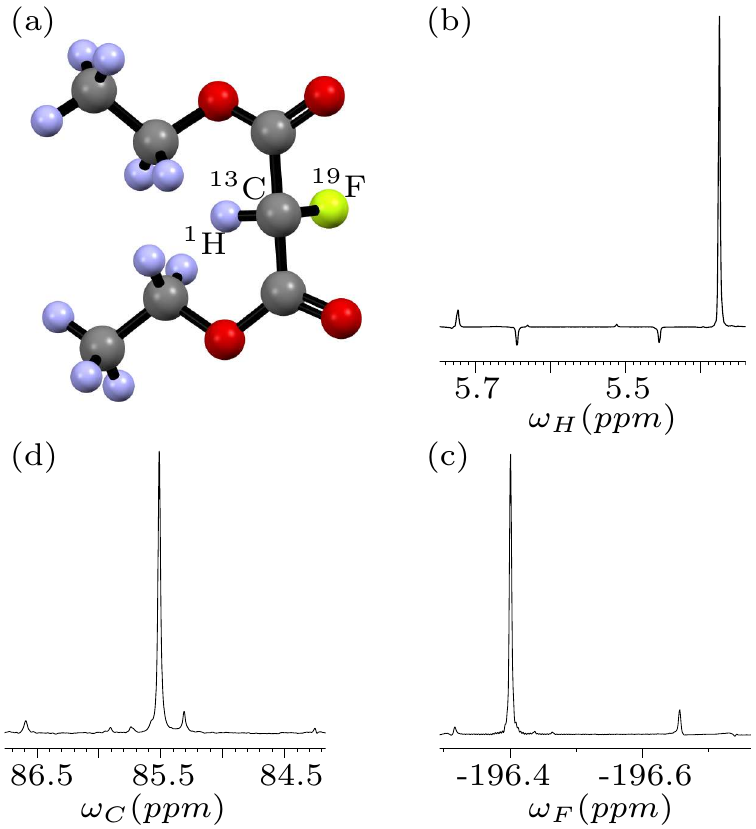} 
\caption{
(a) Molecular structure of ${}^{13}$C-labeled diethyl fluoromalonate used as
the three-qubit quantum system. 
The spectra shown in (b), (c) and (d) correspond to
the $^1$H, $^{19}$F and $^{13}$C 
spins respectively, obtained after applying a $90^{\circ}$
readout pulse on the $\vert 000 \rangle$ pseudopure state. The $J$-couplings
between different nuclei are: $J_{HF} = 47.5 $~Hz, $J_{HC} = 161.5 $~Hz
and $J_{FC} = -191.7 $~Hz. The  spin-lattice relaxation times measured for
different nuclei are: $T^{H}_1 = 3.0 \pm 0.34$~s, $T^{F}_1 = 3.3 \pm 0.15$~s and
$T^{C}_1 = 3.2 \pm 0.38$~s, while the spin-spin relaxation times measured for
different nuclei are: $T^{H}_2 = 1.3 \pm 0.24$~s, $T^{F}_2 = 1.4 \pm 0.22$~s and
$T^{C}_2 = 1.2 \pm 0.18$~s.
} 
\label{fig1}
\end{figure}
\begin{figure*}[t] 
\includegraphics[angle=0,scale=1.1]{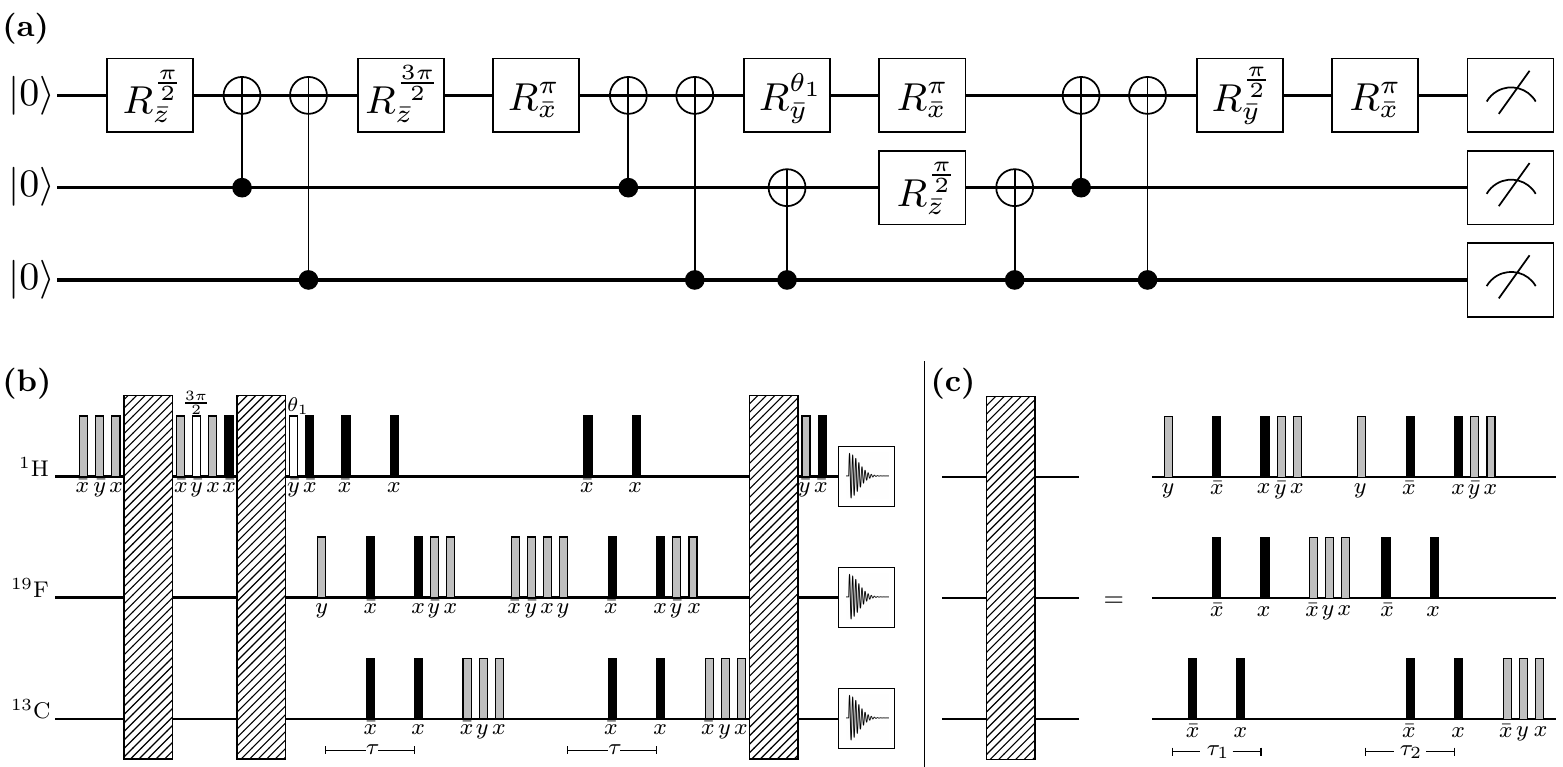} 
\caption{
(a) Quantum circuit to simulate the action of the Kraus operator $A_1$ 
of the phase damping channel on the
initial state $\vert 00 \rangle$ using the Sz.-Nagy algorithm.  The unitary
dilation operator $U_{A_1}$ is realized using eight CNOT gates 
and eight single-qubit
rotation gates $R_{\phi}^{\theta}$.  (b) NMR implementation of the quantum
circuit given in (a). Gray and black filled rectangles 
represent $\pi/2$ and $\pi$ pulses, respectively. The angles of the pulses
represented by unfilled rectangles are shown above each pulse, where
$\theta_1 = 0.3737*\frac{\pi}{2}$. The dashed rectangular blocks consist of
a set of pulses which have been expanded and represented 
in panel (c). The phases
are written below the corresponding pulse.  The free evolution time
periods are set to $\tau = 0.0078$~s, $\tau_{1}=0.0105$~s 
and $\tau_2=0.0031$~s, respectively.
The measurement box is indicated by a decaying time-domain NMR signal.  } 
\label{fig2}
\end{figure*}
\section{Time evolution of open quantum systems and the Sz.-Nagy algorithm}
\label{sec2} 
The Sz.-Nagy algorithm allows the density matrix of the system to
evolve from its initial density matrix $\rho$ to a density matrix $\rho(t)$ at
time $t$, under a given evolution model. To mathematically represent the
framework of the Sz.-Nagy algorithm, consider the operator-sum representation
form for the time evolution of the density matrix (also called the Kraus
operator representation)~\cite{kraus-book-1983}: \begin{equation} \rho(t) =
\sum_i A_i \rho A_i^{\dagger} \label{eq1} \end{equation} where the $A_i$s are
Kraus operators satisfying $\sum_iA_i^{\dagger}A_i=I$.  In order to implement
the Sz.-Nagy algorithm to simulate any given open quantum dynamics, one requires
the complete set of corresponding Kraus operators given in Eq.\ref{eq1}.  

The Sz.-Nagy algorithm states that, for any contraction 
operator $W$ 
acting on a vector $v$
in a
Hilbert space ${\cal H}_1$, 
one can construct a corresponding unitary
dilation unitary operator $U_w$ in a larger Hilbert space ${\cal H}_2$, such
that~\cite{Hu-sr-2020}:
\begin{equation} W^m = P_{{\cal H}_1} U_w^m P_{{\cal H}_1}, \quad  m \leq
N \label{eq4} \end{equation} 
where $P_{{\cal H}_1}$ is the projection
operator
which projects the output vector into
the space ${\cal H}_1$, dim($ {\cal H}_2$)$>$ dim($ {\cal H}_1 $), 
and $m$ and $N$ are integers.
Note that the operator $W$ is said to be
a `contraction' if it preserves or shrinks the norm of any
vector i.e. $\Vert W \Vert = \text{sup} \frac{\Vert Wv \Vert
}{\Vert v \Vert} \leq 1$. 
Eq.\ref{eq4} implies that the
action of the contraction $W$, applied up to $N$ times in space
${\cal H}_1$ can be simulated via the corresponding unitary dilation $
U_w$ applied up to $N$ times on space ${\cal H}_2$, given that the
input vector lies in ${\cal H}_1$ and the output vector is projected
into ${\cal H}_1$.

Consider the set 
of Kraus operators $\lbrace A_i \rbrace$ (Eq.~\ref{eq1}),
corresponding to a given quantum
process, which evolve the initial density matrix to $\rho(t)$.
The Kraus
operator $A$ has to be a `contraction operator',
in order to find its corresponding 
unitary dilation operator.
For a general proof 
that any Kraus operator satisfies all properties 
to be a `contraction operator' see
Reference~\cite{Hu-sr-2020}).
For simplicity, consider an $n$-qubit system with a
corresponding
Hilbert space $\mathcal{H}$ of dimension $2^n$, and let
the initial density matrix $\rho$ be in a pure state i.e. $\rho =
\vert \phi \rangle \langle \phi \vert$. In this case,
the steps to implement the Sz.-Nagy algorithm to simulate
Eq.\ref{eq1}, are as follows~\cite{Hu-sr-2020}: 
\begin{enumerate}
\item Prepare the pure input state $ \vert \Phi \rangle  =
\vert 0 \rangle \otimes \vert \phi \rangle$ in a larger
Hilbert space of dimension $2^{n+1}$ with the help of one
ancillary qubit.  
\item Apply the unitary operation $U_{A_i}$ 
operation on the input state $\vert \Phi \rangle$, where
$U_{A_i}$ is the minimal unitary dilation of $A_i$ (with
$N=1$) given by:
\begin{equation}
{U}_{{A_i}}=\left(\begin{array}{cc} {A}_i &
{D}_{A_i^{\dagger}} \\ {D}_{{A_i}} & -A_i^{\dagger}
\end{array}\right) \label{eq5} \end{equation} where
${D}_{{A_i}} = \sqrt{I-A_i^{\dagger}A_i}$~\cite{Hu-sr-2020}.  
\item 
Project the output vector
$U_{A_i}\vert \Phi \rangle$ into a smaller Hilbert space
$\mathcal{H}$, using the appropriate projection operator
$P_{\mathcal{H}}$, the dimension of $\mathcal{H}$ being $2^n$.  
\item Repeat the above steps for the remaining
Kraus operators, and 
sum over all output density matrices
obtained after
Step~3, in order  
to compute the effect of the given quantum process on the input state
$\rho$.  
\end{enumerate}

Note that if the initial density matrix is in a mixed state 
i.e.  $\rho = \sum_j p_j \vert \phi_j \rangle \langle \phi_j
\vert$, then one has to the repeat Sz.-Nagy algorithm for all
$\vert \phi_j \rangle $, in order to obtain the effect of 
a given quantum
process on the initial mixed-state density matrix.
   
\section{Experimentally simulating two-qubit 
non-unitary quantum processes}
\label{sec3}
We now proceed towards experimentally implementing the Sz.-Nagy algorithm in
order to simulate a two-qubit pure phase damping channel 
and an MFGP process on an NMR quantum
information processor, with the help of one ancillary qubit.  
\subsection{Experimental details}
\label{expt}
We used ${}^{13}$C-labeled diethyl fluoromalonate dissolved
in an acetone-D6 as the three-qubit system, and assigned the
$^1H$, $^{19}F$ and $^{13}C$ spins as the first, second and
third qubit, respectively (see Fig.~\ref{fig1} for
experimental parameters). The Hamiltonian for a system of
three spin-1/2 nuclei in the rotating frame is given by:
\begin{equation} H = -\sum_{i=1}^3 \omega_iI_{iz}+ \sum_{i,j
= 1,i>j}^3 J_{ij}I_{iz}I_{jz} \end{equation} where
$\omega_i$ is the chemical shift of the $i$th spin, $J_{ij}$
is the scalar coupling strength between the $i$th and $j$th
spins and $I_{iz}$ represents the $z$-component of the spin
angular momentum of the $i$th nucleus.  State initialization
was achieved by preparing a pseudopure state (PPS)
corresponding to $\vert 000 \rangle$ from the thermal state
the using spatial averaging
technique~\cite{oliveira-book-07}. The density matrix
$\rho_{000}$ corresponding to $\vert 000\rangle$ PPS is
given by: \begin{equation} \rho_{000}=(\frac{1-\epsilon}{8})
I_8 + \epsilon \vert 000 \rangle\langle 000\vert \label{eq8}
\end{equation} where $\epsilon \approx 10^{-5}$ denotes the
bulk magnetization of the spin ensemble at room temperature
and $I_8$ is the $8 \times 8$ identity matrix.

We performed convex optimization based quantum process tomography
to reconstruct the experimental process matrix that characterizes the
given quantum process. The
process and state fidelity is calculated using the measure ${\mathcal
F}(\chi^{}_{\rm expt},\chi^{}_{\rm theo})$~\cite{Gaikwad-qip-2021}:
\begin{equation} {\mathcal F}(\chi^{}_{\rm expt},\chi^{}_{\rm theo})=
\frac{|{\rm Tr}[\chi^{}_{\rm expt}\chi_{\rm theo}^\dagger]|} {\sqrt{{\rm
Tr}[\chi_{\rm expt}^\dagger\chi^{}_{\rm expt}] {\rm Tr}[\chi_{\rm
theo}^\dagger\chi^{}_{\rm theo}]}} \label{eq10} \end{equation} where
$\chi^{}_{\rm expt}$ ($\rho^{}_{\rm expt}$) and $\chi^{}_{\rm theo}$
($\rho^{}_{\rm theo}$) define the experimental and theoretical process (density)
matrices, respectively.

\subsection{Simulating a two-qubit phase damping channel}
\label{pd}
The phase damping
channel is well known and plays an important role in solution NMR, where it is
responsible for the transverse relaxation of the spin ensemble. In some
real-life situations, the low experimental fidelity of certain quantum gates
(with long implementation times) can be ascribed to the deleterious effects of
the phase damping channel.  Several studies have focused on protecting fragile
quantum coherences in the presence of phase
damping\cite{singh-epl-2017,singh-pra-2017,singh-pra-2018,singh-pramana-2020}.

We use the superoperator form to describe the open
quantum dynamics of a system evolving under the action of  a phase
damping channel, where the generator of the phase damping process is
available~\cite{childs-pra-2001}. 
Let $\mathcal{Z}_1$ and $\mathcal{Z}_2$ denote the
generators corresponding to the phase damping channel acting
independently on qubit 1 and qubit 2, respectively.
The matrix form of the generators $\mathcal{Z}_1$ and
$\mathcal{Z}_2$ is given 
by~\cite{childs-pra-2001,singh-pramana-2020}:
\begin{equation} \begin{aligned} \mathcal{Z}_1=&
\operatorname{diag}\left[0,0,-\gamma_{1},-\gamma_{1},0,
0,-\gamma_{1},-\gamma_{1},-\gamma_{1},-\gamma_{1},0,0,\right.\\
&\left.-\gamma_{1},-\gamma_{1},0, 0\right] \nonumber \\
\mathcal{Z}_2=&
\operatorname{diag}\left[0,-\gamma_{2},0,-\gamma_{2},-\gamma_{2},0
-\gamma_{2},0,0,-\gamma_{2},0-\gamma_{2},\right.\\
&\left.-\gamma_{2},0,-\gamma_{2},0\right] \\ \end{aligned}
\end{equation} 
where $\gamma_1$ and $\gamma_2$ are the phase
damping rates for qubit 1 and qubit 2, respectively. The
resultant process is denoted by the superoperator $\Xi$ which
consists of the simultaneous action of phase damping channel
independently acting on qubit 1 and qubit 2 and has
the generator $\mathcal{Z} = \mathcal{Z}_1+\mathcal{Z}_2$. 
The time evolution of the initial two-qubit density
matrix $\rho$ can be written as~\cite{childs-pra-2001}:
\begin{equation}
\rho(t)=\Xi(\rho)=e^{\mathcal{Z}t}(\vec{\rho}) \label{eq6}
\end{equation} 
In order to simulate Eq.~\ref{eq6} using
the Sz.-Nagy algorithm, one requires the complete set of Kraus
operators corresponding to the phase damping process. We 
used the standard quantum process tomography (QPT)
technique to compute Kraus operators as follows:
\begin{enumerate} 
\item Construct
the complete set of linearly independent initial input
density matrices.  
\item Estimate output density
matrices by evolving each input density matrix using
Eq.~\ref{eq6}.  
\item From knowledge of the input and
output density matrices, compute the process matrix $\chi$ using
the standard QPT protocol.  
\item Using unitary diagonalization of
$\chi$ matrix as~\cite{Gaikwad-qip-2021}: $\chi = VDV^{\dagger}$, compute the complete
set of Kraus operators as: 
\begin{equation}
A_{i}=\sqrt{d_{i}} \sum_{j} V_{j i} E_{j} \label{eq7}
\end{equation} where $A_i$s are the Kraus operators, $d_i$s
are diagonal elements of the matrix $D$, $V_{ji}$s are elements
of the matrix $V$ and the $E_j$s form a fixed operator basis. The
diagonal elements of matrix $D$ are eigenvalues of the $\chi$
matrix and the columns of matrix $V$ are the corresponding
normalized eigenvectors of the $\chi$ matrix.  
\end{enumerate}
Note that the Kraus operators corresponding to the two-qubit
phase damping channel are already known in literature and
one could have directly use them here. However we have used the
superoperator form given in Eq.~(\ref{eq6}), in order to 
illustrate our method  which is general and can be used to
describe
quantum processes where the Kraus operators
are not directly available (such as correlated phase damping
channels~\cite{childs-pra-2001}). 
The unitary dilation operators $\lbrace
U_{A_i} \rbrace$ corresponding to each Kraus operator
$\lbrace A_i \rbrace $ were computed using Eq.~\ref{eq5}. 
We set the values of the phase damping rates to be
$\gamma_1=1.4$ and $\gamma_2=1.5$, and evolved the initial
density matrix for a time $t=2$ s using Eq.~\ref{eq6}. 
We note here that the time required to implement a unitary dilation operator on
two NMR qubits depends crucially on the implementation times of the CNOT gates,
which for our system turns out to be in the range of 3-11 ms. The total time
required to implement all the four unitary dilation operators required to
simulate the phase damping channel is hence $\approx 80$ ms.  The spin-spin
relaxation times (T$_2$) of the three NMR qubits (which characterizes the
natural phase damping channel active in the NMR system) are: T$_{2}^{H}=1.3$~s,
T$_{2}^{F}=1.4$~s, T$_{2}^{C}=1.2$~s, respectively. Since the time required to
implement the unitary dilation operators is much smaller than the natural phase
damping rates of the system, the experimental implementation of the simulated
phase damping channel is largely unaffected by the natural NMR noise.

The
complete set of Kraus operators which evolve the initial
density matrix under the action of independent phase damping
channels on each qubit, for given values of $\gamma_1$,
$\gamma_2$ and $t$, is given in Appendix~\ref{appendixA}.
It turns out that there are four non-zero Kraus operators
which characterize the phase damping channel for the given
parameter values.  

Fig.~\ref{fig2} demonstrates the implementation of the Sz.-Nagy algorithm to
simulate the action of Kraus operator $A_1$ (see Appendix~\ref{appendixA}) on
the two-qubit initial input state $\vert \phi \rangle \langle \phi \vert = \vert
00 \rangle \langle 00 \vert$.  The initial two-qubit state $\vert 00 \rangle $
is encoded in a three-qubit input state as $\vert 000 \rangle = \vert 0
\rangle_a \otimes \vert 00 \rangle_{\rm{main}}$.  The quantum circuit given in
Fig.\ref{fig2} (a) represents the action of the unitary dilation operator
$U_{A_1}$ on the input state $\rho_{000}$, followed by measurement. We used the
column-by-column decomposition (COC) method~\cite{iten-pra-2016,
iten-arxiv-2021} to decompose three-qubit unitary dilation operators $\lbrace
U_{A_i} \rbrace $. Using the COC  method, $U_{A_1}$ is realized using eight CNOT
gates and eight single-qubit rotation gates $R_{\phi}^{\theta}$ (where $\phi$
denotes the axis of rotation and $\theta$ denotes the angle of rotation).  The
COC decompositions of the other unitary dilation operators are given in
Appendix~\ref{appendixA}.  We note here in passing that the same quantum circuit
one can also be used to simulate the action of $A_1$ on arbitrary initial
two-qubit states $\rho = \vert \phi \rangle \langle \phi \vert$, in which case
we merely need to prepare the three-qubit system in the state $\vert 0 \rangle
\otimes \vert \phi \rangle$. Further, the arbitrary initial input state $\vert
\phi \rangle$ of the two-qubit system lies in the smaller Hilbert space which is
spanned by the vectors: $\vert 000 \rangle $, $\vert 001 \rangle $, $\vert 010
\rangle $ and $\vert 011 \rangle $. The action of projecting the
higher-dimensional output state into this smaller Hilbert space is equivalent to
estimating a $4 \times 4$ dimensional partial density matrix  (corresponding to
the first four rows and columns of the higher-dimensional output density
matrix).  The NMR pulse sequence to implement the quantum circuit is depicted in
Fig.\ref{fig2}(b).  Spin-selective high-power rf pulses were used to implement
single-qubit rotation gates.  Filled gray and black rectangles in
Fig.~\ref{fig2}(b) represent $\pi/2$ and $\pi$ pulses respectively, while
unfilled rectangles represent pulses with their corresponding flip angles given
above each pulse; the value of $\theta_1$ was set to $0.3737*\frac{\pi}{2}$. The
three dashed boxes consist of a set of pulses which have been expanded and
depicted in Fig.~\ref{fig2}(c). The phase of each pulse is shown below every
rectangle. The various free evolution time periods were set to $\tau = 0.0078$
s, $\tau_{1}=0.0105$ s and $\tau_2=0.0031$ s, respectively.  The measurement box
depict the decaying time domain NMR signal (the free induction decay (FID))
which is Fourier transformed to obtain the NMR spectrum.  Finally, tomographic
measurements were performed to compute density matrix elements $\lbrace
\rho_{ij}$, $0 \leq i,j \leq 4 \rbrace$.  The normalized trace distance between
the experimentally obtained output Hermitian matrix $(A_1 \rho_{00}
A_1^{\dagger})_{{\rm exp}}$ and the theoretically expected matrix $(A_1
\rho_{00} A_1^{\dagger})_{{\rm the}}$ turns out to be $0.9885$.  A similar
quantum circuit and NMR pulse sequence is employed to simulate the MFGP,
where the action of the Kraus
operator $A_1$ (Appendix~\ref{appendixB}) 
on the state $\vert 00 \rangle$ can be
simulated using the unitary dilation operator $U$ (Eq.~\ref{c1} in
Appendix~\ref{appendixB}), via the COC method 
and 9 CNOT gates and 18 local 
rotations.

\begin{table}[h!] 
\caption{\label{table1} The normalized
trace distance between the experimentally obtained output
Hermitian matrix $(A_i \rho_j A_i^{\dagger})_{{\rm exp}}$
and the theoretically expected matrix $(A_i \rho_j
A_i^{\dagger})_{{\rm the}}$ for the phase damping channel.}
\renewcommand{\arraystretch}{1.3}
\begin{tabular}{c | c | c | c | c}
\hline \hline ~~~~~&~~~~~ $A_1$ ~~~~~& ~~~~~~$A_2$~~~~~~&~~~~~ $A_3$~~~~~~&~~~~~~$A_4$~~~~~\\
\hline \hline $\vert 00 \rangle $ & 0.9885 &   0.9881 &  0.9911   & 0.9773 \\
$\vert 01 \rangle $ & 0.9769  &  0.9909  &  0.9901  &  0.9804 \\
$\vert 0+ \rangle $ & 0.9105  &  0.9825  &  0.9877  &  0.9788 \\
$\vert 0- \rangle $ & 0.8984  &  0.9905  &  0.9585  &  0.9613 \\
$\vert 10 \rangle $ & 0.9625  &  0.9579  &  0.9873  &  0.9780 \\
$\vert 11 \rangle $ & 0.9071  &  0.9519  &  0.9847  &  0.9747 \\
$\vert 1+ \rangle $ & 0.8193  &  0.9447  &  0.9538  &  0.9823 \\
$\vert 1- \rangle $ & 0.7236  &  0.9212  &  0.9286  &  0.9494 \\
$\vert +0 \rangle $ & 0.9625  &  0.9888  &  0.9251  &  0.9608 \\
$\vert +1 \rangle $ & 0.8733  &  0.9447  &  0.9892  &  0.9856 \\
$\vert ++ \rangle $ & 0.8697  &  0.9590  &  0.9712  &  0.9814 \\
$\vert +- \rangle $ & 0.8974  &  0.9489  &  0.9506  &  0.9669 \\
$\vert -0 \rangle $ & 0.9149  &  0.9892  &  0.9205  &  0.9630 \\
$\vert -1 \rangle $ & 0.8168  &  0.9222  &  0.9946  &  0.9781 \\
$\vert -+ \rangle $ & 0.8105  &  0.9363  &  0.9819  &  0.9643 \\
$\vert -- \rangle $ & 0.8233  &  0.9518  &  0.9608  &  0.9594 \\
\hline \end{tabular}
\end{table}
\begin{figure}[t]
\includegraphics[angle=0,scale=1]{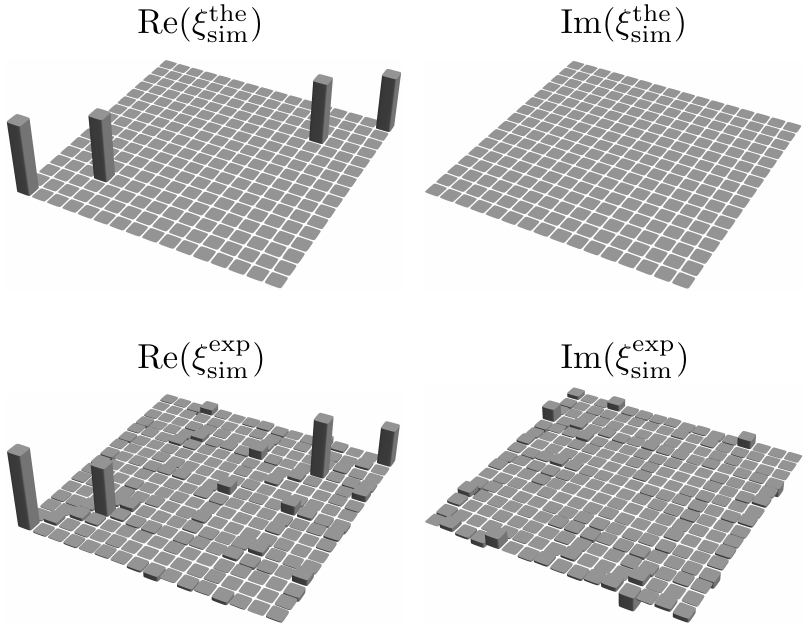} 
\caption{Process matrices obtained by theoretically and
experimentally simulating phase damping channels acting
independently on each qubit in a two-qubit NMR system.  
The bar plots in the first
column represent the real part of the  process matrices of
the theoretically simulated 
($\rm{Re}(\chi_{\rm{theo}}))$, and the experimentally
obtained phase damping channel computed using convex
optimization-based QPT ($\rm{Re}(\chi_{\rm{expt}}))$, respectively.   
The bar plots in the second column represent
the imaginary part of the respective process matrices.  } 
\label{fig3} 
\end{figure}

For the phase damping channel,
the normalized trace
distance between the experimentally obtained output Hermitian
matrix ($A_i \rho_j A_i^{\dagger}$) and the theoretically
expected matrix is
given in Table~\ref{table1}. 
High values of the normalized trace distance in
Table~\ref{table1} clearly demonstrates
the successful experimental simulation of 
the action of a given Kraus operator on
a given initial state. However, for all the initial
quantum states, the values given in the first column corresponding to $A_1$ are
relatively smaller than the values given in the other columns
corresponding to $A_2$, $A_3$ and $A_4$, respectively. This is 
due to the fact that the
experimental complexity involved in implementing 
$U_{A_1}$ is relatively larger than $U_{A_2}$, $U_{A_3}$ and $U_{A_4}$ in terms
of the number of CNOT gates which introduces more experimental errors in
the implementation of
$U_{A_1}$. In addition,
it turns out that for some quantum states such as $\vert 1+
\rangle$, $\vert 1- \rangle$, $\vert -1 \rangle$, $\vert -+ \rangle$ and $\vert
-- \rangle$ the values of trace distance are relatively small as compared to
other states, which can be attributed to errors 
in initial state preparation.

Note that in order to obtain
the final density matrix $\rho(t)$ evolved under a desired quantum
channel one has to assemble all results from each
Kraus operator. For the phase damping channel 
it turns out that four quantum circuits
corresponding
to each $U_{A_i}$ need to be implemented to obtain the final
$\rho(t)$. 
For completeness, 
we experimentally simulated the
action of all Kraus operators corresponding to the
phase damping channel on 16 linearly
independent two-qubit density matrices. 
For all 16 states, the fidelities between the
experimentally simulated state using the Sz.-Nagy algorithm 
and the theoretically simulated
state  for the phase damping channel are
given in 
Table~\ref{table3}.
\begin{table}[h] 
\caption{\label{table3} Fidelity between the experimentally
and theoretically simulated two-qubit states evolving under
independent phase damping channels.}
\renewcommand{\arraystretch}{1.3}
\begin{tabular}{c| c| c| c}
\hline \hline State &
~~~Fidelity~~~& ~~~State~~~& ~~~Fidelity~~~\\
\hline \hline $\vert 00 \rangle $ & 0.9936 & $\vert +0 \rangle $ & 0.9568 \\ 
$\vert 01 \rangle $ & 0.9950 & $\vert +1 \rangle $ & 0.9564\\ 
$\vert 0+ \rangle $ & 0.9734 & $\vert ++ \rangle $ & 0.9673 \\ 
$\vert 0- \rangle $ & 0.9696 & $\vert +- \rangle $ & 0.9607 \\ 
$\vert 10 \rangle $ & 0.9885 & $\vert -0 \rangle $ & 0.9444\\ 
$\vert 11 \rangle $ & 0.9821 & $\vert -1 \rangle $ & 0.9343\\ 
$\vert 1+ \rangle $ & 0.9521 & $\vert -+ \rangle $ & 0.9412\\ 
$\vert 1- \rangle $ & 0.9409 & $\vert -- \rangle $ & 0.9373\\ 
\hline \end{tabular}
\end{table}

The high values of the fidelities given in Table~\ref{table3} demonstrate the
successful experimental simulation of the action of the phase damping channel on
a given initial quantum state.  Since the given set of 16 states forms a
complete basis set, one can simulate the action of the phase damping channel on
an arbitrary quantum state with fidelities ranging between 0.9343 and 0.9950. 

\subsection{Simulating a magnetic field gradient pulse}
\label{mfgp}
MFGP are extensively used in NMR and 
magnetic resonance imaging experiments, covering a wide 
range of applications, such as studies of molecular
diffusion and spatial encoding for 
imaging~\cite{denis-neuro-2012,paga-analyst-2017,han-chemmat-2021}.  
Recently, a time and
space discretization method was proposed to simulate shaped gradient
pulses~\cite{laflamme-arxiv-2020}.  The action of a MFGP is similar to the phase
damping channel, as it effectively kills the off-diagonal elements (coherences)
of the density matrix in a controlled manner. In this study we employ the
Sz.-Nagy algorithm to simulate two-qubit dynamics under the action of a shaped
MFGP applied for a given duration.

A shaped MFGP has a strength that varies during its execution. The gradient
pulse is defined by a list of values, with each element in the list defining the
relative gradient strength during a particular time interval.  The interval
length is defined by the length of the entire gradient shape divided by the
number of intervals. The gradient strength is expressed as a percentage of the
maximum strength. In the NMR hardware, MFGP is applied using gradient coils.
The parameters of the shaped gradient pulse used are: Sine shaped, duration
1000$\mu$s, number of time intervals =100, and an applied gradient strength of
$15\%$.

In order to simulate the desired  MFGP using the Sz.-Nagy algorithm, we first
need to characterize it and then compute the corresponding Kraus operators. We
used convex optimization based quantum process tomography to experimentally
characterize the desired MFGP and then computed the complete set of Kraus
operator using Eq.~(\ref{eq7}).  To achieve this, we experimentally prepared the
complete set of linearly independent initial two-qubit quantum states: $\lbrace
\vert 0\rangle, \vert 1\rangle, \vert +\rangle ,\vert -\rangle \rbrace^{\otimes
2} $ where $\vert + \rangle = (\vert 0 \rangle + \vert 1 \rangle)/\sqrt{2}$ and
$\vert - \rangle = (\vert 0 \rangle + i\vert 1 \rangle)/\sqrt{2}$.  The desired
MFGP is then applied on the initial input states using gradient coils.  By
performing full quantum state tomography of all output states we compute the
process matrix $\chi$ characterizing the MFGP  and the complete set of Kraus
operators are calculated using Eq.~(\ref{eq7}).  The Sz.-Nagy algorithm is
finally employed to simulate the MFGP using only unitary operations. At the end,
the process fidelity is computed between the experimental process matrix
characterizing the MFGP and the experimental process matrix of the simulated
MFGP. The complete set of Kraus operators and corresponding unitary dilation
operators for the shaped MFGP are given in Appendix\ref{appendixB}.

\begin{table}[h!] 
\caption{\label{table2} The normalized trace distance between experimentally
simulated output hermitian matrix $(A_i \rho_j
A_i^{\dagger})^{\rm{exp}}_{\rm{sim}}$ using SND algorithm and experimentally
obtained matrix $(A_i \rho_j A_i^{\dagger})^{\rm{exp}}_{\rm{qpt}}$ via 
quantum process tomography of MFGP implemented on two qubits.}
\renewcommand{\arraystretch}{1.3} 
\begin{tabular}{c | c | c | c | c}
\hline \hline ~~~~~&~~~~~ $A_1$ ~~~~~& ~~~~~~$A_2$~~~~~~&~~~~~ $A_3$~~~~~~&~~~~~~$A_4$~~~~~\\
\hline \hline
$\vert 00 \rangle $ & 0.8671 &   0.9541  &  0.9704  &  0.9182 \\
$\vert 01 \rangle $ &0.8919  &  0.9866   & 0.9602   & 0.9884 \\
$\vert 0+ \rangle $ &0.8705  &  0.8359  &  0.9164  &  0.9476  \\
$\vert 0- \rangle $ &0.9204  &  0.8132  &  0.9208  &  0.8494  \\
$\vert 10 \rangle $ &0.9925  &  0.8317  &  0.9225  &  0.9685  \\
$\vert 11 \rangle $ &0.9732  &  0.9627  &  0.9805  &  0.9345 \\
$\vert 1+ \rangle $ &0.9573  &  0.7373  &  0.9715  &  0.7062  \\
$\vert 1- \rangle $ &0.9025  &  0.8016  &  0.9654  &  0.6549  \\
$\vert +0 \rangle $ &0.7818  &  0.6658  &  0.9220  &  0.9097  \\
$\vert +1 \rangle $ &0.8430  &  0.9697  &  0.8851  &  0.6822  \\
$\vert ++ \rangle $ &0.7669  &  0.7404  &  0.8063  &  0.7433  \\
$\vert +- \rangle $ &0.8397  &  0.7700  &  0.7414  &  0.6495 \\
$\vert -0 \rangle $ &0.7312  &  0.7444  &  0.9365  &  0.9148 \\
$\vert -1 \rangle $ &0.8902  &  0.9767  &  0.8799  &  0.6585  \\
$\vert -+ \rangle $ &0.8247  &  0.8123  &  0.8282  &  0.7211  \\
$\vert -- \rangle $ &0.8410  &  0.8181  &  0.8395  &  0.6690  \\
\hline \end{tabular}
\end{table}

\begin{table}[h] 
\caption{\label{table4} Fidelity between experimentally 
simulated and experimentally implemented
two-qubit state under 
the action of a magnetic field gradient pulse.}
\renewcommand{\arraystretch}{1.3}
\begin{tabular}{c c c c}
\hline \hline State &
~~~Fidelity~~~& ~~~State~~~& ~~~Fidelity~~~\\
\hline \hline
$\vert 00 \rangle $ & 0.9818 & $\vert +0 \rangle $ & 0.9069  \\
$\vert 01 \rangle $ & 0.9884 & $\vert +1 \rangle $ & 0.9440 \\
$\vert 0+ \rangle $ & 0.9893 & $\vert ++ \rangle $ & 0.9096 \\
$\vert 0- \rangle $ & 0.9619 & $\vert +- \rangle $ & 0.8888 \\
$\vert 10 \rangle $ & 0.9719 & $\vert -0 \rangle $ & 0.8946\\
$\vert 11 \rangle $ & 0.9754 & $\vert -1 \rangle $ & 0.9514 \\
$\vert 1+ \rangle $ & 0.9577 & $\vert -+ \rangle $ & 0.9311 \\
$\vert 1- \rangle $ & 0.9426 & $\vert -- \rangle $ & 0.9097 \\
\hline \end{tabular}
\end{table}

For the shaped MFGP operation, the normalized trace distance
between the simulated output hermitian matrix $(A_i \rho_j
A_i^{\dagger})^{\rm{exp}}_{\rm{sim}}$ using 
the Sz.-Nagy algorithm and the experimentally
obtained matrix $(A_i \rho_j A_i^{\dagger})^{\rm{exp}}_{\rm{qpt}}$  via
quantum process tomography
is given in Table~\ref{table2}.
For the case of the MFGP process,  
the Kraus operators which are to be
experimentally simulated are themselves 
computed from experimentally constructed process matrix and 
also have a relatively high experimental 
complexity in terms of the number of CNOT gates required 
to experimentally implement the unitary dilation
operators.
This could be a possible explanation for the smaller values of the
trace distance in Table~\ref{table2} for the MFGP process, as compared
to the phase damping channel.

For the MFGP operation, four quantum circuits corresponding to each $U_{A_i}$
need to be implemented to obtain the final $\rho(t)$.  For completeness, we
experimentally simulated the action of all Kraus operators on 16 linearly
independent two-qubit density matrices for the MFGP process.  For all 16 states,
the fidelities between the experimentally simulated state using the Sz.-Nagy
algorithm and the theoretically simulated state  corresponding to the MFGP
operation are given in Table~\ref{table4}.  Since the the set of 16 states given
in Table~\ref{table4} forms a complete basis set, the action of the MFGP can be
simulated on any arbitrary quantum state with fidelities ranging between 0.8888
and 0.9893. 

\begin{figure}[h] 
\includegraphics[angle=0,scale=1]{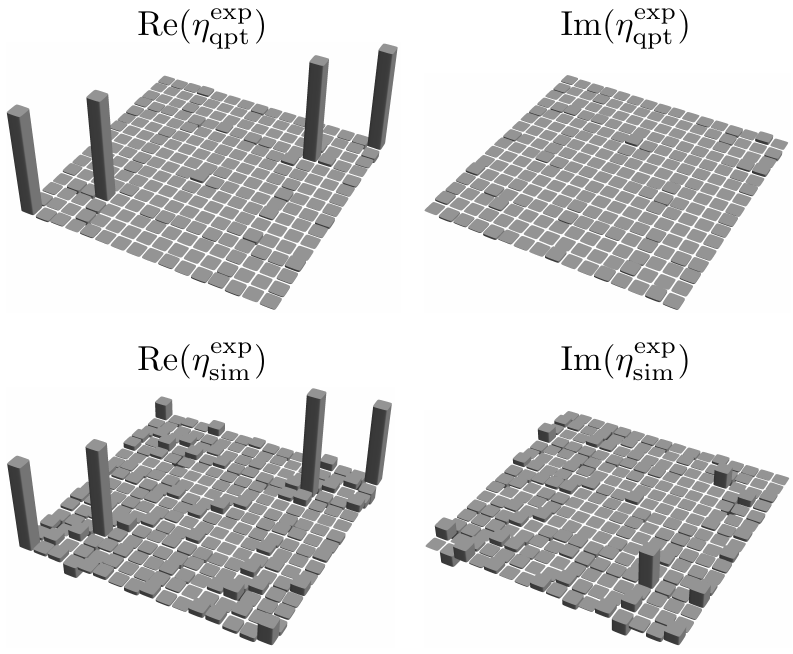} 
\caption{ In
the top panel, $\rm{Re}(\eta^{\rm exp}_{\rm qpt})$ and $\rm{Im}(\eta^{\rm
exp}_{\rm qpt})$ denote the real and imaginary parts of the experimental
process matrix obtained by performing quantum process tomography of the
shaped MFGP applied on two qubits.
In the bottom panel,
$\rm{Re}(\eta^{\rm exp}_{\rm sim})$ and  $\rm{Im}(\eta^{\rm exp}_{\rm sim})$
represent the real and imaginary parts of the process matrix of
the same MFGP 
experimentally simulated using the Sz.-Nagy algorithm.
The process fidelity between $\eta^{\rm exp}_{\rm qpt}$ and $\eta^{\rm exp}_{\rm
sim}$ turns out to be 0.8824.} 
\label{fig4} 
\end{figure}

For the phase damping channel, the process fidelity is computed between
$(\xi^{\rm exp}_{\rm sim})$ obtained via the experimentally simulated channel
and $(\xi^{\rm the}_{\rm sim})$ obtained via the theoretically simulated
channel. For the shaped MFGP process, the process fidelity is computed between
$(\eta^{\rm exp}_{\rm qpt})$ obtained via the experimentally simulated shaped
MFGP and $(\eta^{\rm exp}_{\rm qpt})$ obtained via  quantum process tomography
performed on experimentally implemented shaped MFGP.
For the phase damping channel, the computed process fidelity turns out to be
0.9148 and  the respective tomographs are compared in Fig.~\ref{fig3}, while for
the shaped MFGP process, the computed process fidelity turns out to be 0.8824
and the respective tomographs are compared in Fig.~\ref{fig4}, where the
tomographs are plotted on the same scale.  For both the phase damping channel
and the MFGP process, the real part of the process matrix has only four non-zero
elements corresponding to $E_1 = I \otimes I $, $E_4 = I \otimes \sigma_z $,
$E_{13} = \sigma_z \otimes I $, and $E_{16} = \sigma_z \otimes \sigma_z $, which
are the set of Kraus operators~\cite{nielsen-book-10}.  The imaginary part of
the process matrix for both processes turns out to be almost zero.  
From
Figs.\ref{fig3} and \ref{fig4} it can be seen that the action of the shaped MFGP
and the phase damping channel is similar, in effectively destroying the
off-diagonal elements of the density matrix.  
The
deviations in the simulated process matrix from the desired process matrix
arises  due to experimental errors in state preparation, implementation of
unitary dilation operators and inevitable systematic errors.  These errors can
be reduced using appropriate optimization protocols~\cite{Devra-qip-2018}.  
Particularly for the MFGP process, the experimental implementation of all four
unitary dilation operators requires 9 CNOT gates (\ie  9 CNOT gates $\times$ 4
Kraus operators = 36 CNOT gates in total to simulate 
the MFGP process), while 
for the phase damping channel, the 
experimental implementation of the unitary dilation operator $U_{A_1}$,
$U_{A_2}$, $U_{A_3}$ and $U_{A_4}$ requires 8, 3, 3 and 0 CNOT gates,
respectively (\ie 14 CNOT gates in total to simulate the phase
damping  channel). Hence the 
experimental errors are higher in simulating 
the MFGP process as compared to the phase damping channel, which is
reflected in lower values of the process fidelities in Tables~\ref{table2}
and \ref{table4}.
\section{Conclusions}
\label{sec5}
We experimentally implemented the Sz.-Nagy algorithm to simulate an independent
phase damping channel and a shaped MFGP acting on two qubits, with the help of
one ancilla qubit on an NMR quantum information processor. We designed a
protocol to compute the complete set of Kraus operators using quantum process
tomography and the unitary diagonalization technique. 
To validate the quality of the
experimentally simulated quantum process, we performed quantum process
tomography based on the constrained convex optimization technique.  Our results
demonstrate that the experimental implementation of the Sz.-Nagy algorithm is
viable, since it requires only one ancilla qubit to simulate
arbitrary-dimensional open quantum dynamics.  The protocol is general and valid
for arbitrary quantum processes and can be adapted for other physical platforms
to simulate complex quantum processes using the Sz.-Nagy algorithm.  However,
implementing the unitary dilation operator corresponding to a given Kraus
operator remains a challenging task. The need of the hour is hence to develop
computationally efficient algorithms to decompose a given unitary dilation
operator into a universal set of quantum gates. 
\begin{acknowledgments}
All the experiments were performed on a Bruker Avance-III
600 MHz FT-NMR spectrometer at the NMR Research Facility of
IISER Mohali.
Arvind
acknowledges financial support from
DST/ICPS/QuST/Theme-1/2019/General Project number Q-68.
K.D. acknowledges financial support from
DST/ICPS/QuST/Theme-2/2019/General Project number Q-74.
\end{acknowledgments}

\begin{thebibliography}{41}%
\makeatletter
\providecommand \@ifxundefined [1]{%
 \@ifx{#1\undefined}
}%
\providecommand \@ifnum [1]{%
 \ifnum #1\expandafter \@firstoftwo
 \else \expandafter \@secondoftwo
 \fi
}%
\providecommand \@ifx [1]{%
 \ifx #1\expandafter \@firstoftwo
 \else \expandafter \@secondoftwo
 \fi
}%
\providecommand \natexlab [1]{#1}%
\providecommand \enquote  [1]{``#1''}%
\providecommand \bibnamefont  [1]{#1}%
\providecommand \bibfnamefont [1]{#1}%
\providecommand \citenamefont [1]{#1}%
\providecommand \href@noop [0]{\@secondoftwo}%
\providecommand \href [0]{\begingroup \@sanitize@url \@href}%
\providecommand \@href[1]{\@@startlink{#1}\@@href}%
\providecommand \@@href[1]{\endgroup#1\@@endlink}%
\providecommand \@sanitize@url [0]{\catcode `\\12\catcode `\$12\catcode
  `\&12\catcode `\#12\catcode `\^12\catcode `\_12\catcode `\%12\relax}%
\providecommand \@@startlink[1]{}%
\providecommand \@@endlink[0]{}%
\providecommand \url  [0]{\begingroup\@sanitize@url \@url }%
\providecommand \@url [1]{\endgroup\@href {#1}{\urlprefix }}%
\providecommand \urlprefix  [0]{URL }%
\providecommand \Eprint [0]{\href }%
\providecommand \doibase [0]{http://dx.doi.org/}%
\providecommand \selectlanguage [0]{\@gobble}%
\providecommand \bibinfo  [0]{\@secondoftwo}%
\providecommand \bibfield  [0]{\@secondoftwo}%
\providecommand \translation [1]{[#1]}%
\providecommand \BibitemOpen [0]{}%
\providecommand \bibitemStop [0]{}%
\providecommand \bibitemNoStop [0]{.\EOS\space}%
\providecommand \EOS [0]{\spacefactor3000\relax}%
\providecommand \BibitemShut  [1]{\csname bibitem#1\endcsname}%
\let\auto@bib@innerbib\@empty
\bibitem [{\citenamefont {Feynman}(1982)}]{feynman-ijtp-1982}%
  \BibitemOpen
  \bibfield  {author} {\bibinfo {author} {\bibfnamefont {R.~P.}\ \bibnamefont
  {Feynman}},\ }\href {\doibase 10.1007/BF02650179} {\bibfield  {journal}
  {\bibinfo  {journal} {Int. J. Theor. Phys.}\ }\textbf {\bibinfo {volume}
  {21}},\ \bibinfo {pages} {467} (\bibinfo {year} {1982})}\BibitemShut
  {NoStop}%
\bibitem [{\citenamefont {Lloyd}(1996)}]{lloyd-science-1996}%
  \BibitemOpen
  \bibfield  {author} {\bibinfo {author} {\bibfnamefont {S.}~\bibnamefont
  {Lloyd}},\ }\href {\doibase 10.1126/science.273.5278.1073} {\bibfield
  {journal} {\bibinfo  {journal} {Science}\ }\textbf {\bibinfo {volume}
  {273}},\ \bibinfo {pages} {1073} (\bibinfo {year} {1996})}\BibitemShut
  {NoStop}%
\bibitem [{\citenamefont {Kassal}\ \emph {et~al.}(2011)\citenamefont {Kassal},
  \citenamefont {Whitfield}, \citenamefont {Perdomo-Ortiz}, \citenamefont
  {Yung},\ and\ \citenamefont {Aspuru-Guzik}}]{kassal-arpc-2011}%
  \BibitemOpen
  \bibfield  {author} {\bibinfo {author} {\bibfnamefont {I.}~\bibnamefont
  {Kassal}}, \bibinfo {author} {\bibfnamefont {J.~D.}\ \bibnamefont
  {Whitfield}}, \bibinfo {author} {\bibfnamefont {A.}~\bibnamefont
  {Perdomo-Ortiz}}, \bibinfo {author} {\bibfnamefont {M.-H.}\ \bibnamefont
  {Yung}}, \ and\ \bibinfo {author} {\bibfnamefont {A.}~\bibnamefont
  {Aspuru-Guzik}},\ }\href {\doibase 10.1146/annurev-physchem-032210-103512}
  {\bibfield  {journal} {\bibinfo  {journal} {Annu. Rev. Phys. Chem.}\ }\textbf
  {\bibinfo {volume} {62}},\ \bibinfo {pages} {185} (\bibinfo {year} {2011})},\
  \bibinfo {note} {pMID: 21166541},\ \Eprint
  {http://arxiv.org/abs/https://doi.org/10.1146/annurev-physchem-032210-103512}
  {https://doi.org/10.1146/annurev-physchem-032210-103512} \BibitemShut
  {NoStop}%
\bibitem [{\citenamefont {Georgescu}\ \emph {et~al.}(2014)\citenamefont
  {Georgescu}, \citenamefont {Ashhab},\ and\ \citenamefont
  {Nori}}]{franco-rmp-2014}%
  \BibitemOpen
  \bibfield  {author} {\bibinfo {author} {\bibfnamefont {I.~M.}\ \bibnamefont
  {Georgescu}}, \bibinfo {author} {\bibfnamefont {S.}~\bibnamefont {Ashhab}}, \
  and\ \bibinfo {author} {\bibfnamefont {F.}~\bibnamefont {Nori}},\ }\href
  {\doibase 10.1103/RevModPhys.86.153} {\bibfield  {journal} {\bibinfo
  {journal} {Rev. Mod. Phys.}\ }\textbf {\bibinfo {volume} {86}},\ \bibinfo
  {pages} {153} (\bibinfo {year} {2014})}\BibitemShut {NoStop}%
\bibitem [{\citenamefont {Nielsen}\ and\ \citenamefont
  {Chuang}(2010)}]{nielsen-book-10}%
  \BibitemOpen
  \bibfield  {author} {\bibinfo {author} {\bibfnamefont {M.~A.}\ \bibnamefont
  {Nielsen}}\ and\ \bibinfo {author} {\bibfnamefont {I.~L.}\ \bibnamefont
  {Chuang}},\ }\href@noop {} {\emph {\bibinfo {title} {Quantum Computation and
  Quantum Information}}}\ (\bibinfo  {publisher} {Cambridge University Press},\
  \bibinfo {address} {Cambridge UK},\ \bibinfo {year} {2010})\BibitemShut
  {NoStop}%
\bibitem [{\citenamefont {DiVincenzo}(2000)}]{divi-fdp-2000}%
  \BibitemOpen
  \bibfield  {author} {\bibinfo {author} {\bibfnamefont {D.~P.}\ \bibnamefont
  {DiVincenzo}},\ }\href {\doibase
  10.1002/1521-3978(200009)48:9/11<771::AID-PROP771>3.0.CO;2-E} {\bibfield
  {journal} {\bibinfo  {journal} {Fortschritte der Physik}\ }\textbf {\bibinfo
  {volume} {48}},\ \bibinfo {pages} {771} (\bibinfo {year} {2000})}\BibitemShut
  {NoStop}%
\bibitem [{\citenamefont {Harper}\ \emph {et~al.}(2020)\citenamefont {Harper},
  \citenamefont {Flammia},\ and\ \citenamefont {Wallman}}]{harper-np-2020}%
  \BibitemOpen
  \bibfield  {author} {\bibinfo {author} {\bibfnamefont {R.}~\bibnamefont
  {Harper}}, \bibinfo {author} {\bibfnamefont {S.~T.}\ \bibnamefont {Flammia}},
  \ and\ \bibinfo {author} {\bibfnamefont {J.~J.}\ \bibnamefont {Wallman}},\
  }\href {\doibase 10.1038/s41567-020-0992-8} {\bibfield  {journal} {\bibinfo
  {journal} {Nat. Phys.}\ }\textbf {\bibinfo {volume} {16}},\ \bibinfo {pages}
  {1184} (\bibinfo {year} {2020})}\BibitemShut {NoStop}%
\bibitem [{\citenamefont {Breuer}\ and\ \citenamefont
  {Petruccione}(2007)}]{Breuer2007}%
  \BibitemOpen
  \bibfield  {author} {\bibinfo {author} {\bibfnamefont {H.-P.}\ \bibnamefont
  {Breuer}}\ and\ \bibinfo {author} {\bibfnamefont {F.}~\bibnamefont
  {Petruccione}},\ }\href {\doibase 10.1093/acprof:oso/9780199213900.001.0001}
  {\emph {\bibinfo {title} {The Theory of Open Quantum Systems}}}\ (\bibinfo
  {publisher} {Oxford University Press},\ \bibinfo {address} {Oxford},\
  \bibinfo {year} {2007})\ p.\ \bibinfo {pages} {656}\BibitemShut {NoStop}%
\bibitem [{\citenamefont {Rotter}\ and\ \citenamefont
  {Bird}(2015)}]{Rotter-rpp-2015}%
  \BibitemOpen
  \bibfield  {author} {\bibinfo {author} {\bibfnamefont {I.}~\bibnamefont
  {Rotter}}\ and\ \bibinfo {author} {\bibfnamefont {J.~P.}\ \bibnamefont
  {Bird}},\ }\href {\doibase 10.1088/0034-4885/78/11/114001} {\bibfield
  {journal} {\bibinfo  {journal} {Rep. Prog. Phys.}\ }\textbf {\bibinfo
  {volume} {78}},\ \bibinfo {pages} {114001} (\bibinfo {year}
  {2015})}\BibitemShut {NoStop}%
\bibitem [{\citenamefont {Zuniga-Hansen}\ \emph {et~al.}(2012)\citenamefont
  {Zuniga-Hansen}, \citenamefont {Chi},\ and\ \citenamefont
  {Byrd}}]{zuniga-pra-2012}%
  \BibitemOpen
  \bibfield  {author} {\bibinfo {author} {\bibfnamefont {N.}~\bibnamefont
  {Zuniga-Hansen}}, \bibinfo {author} {\bibfnamefont {Y.-C.}\ \bibnamefont
  {Chi}}, \ and\ \bibinfo {author} {\bibfnamefont {M.~S.}\ \bibnamefont
  {Byrd}},\ }\href {\doibase 10.1103/PhysRevA.86.042335} {\bibfield  {journal}
  {\bibinfo  {journal} {Phys. Rev. A}\ }\textbf {\bibinfo {volume} {86}},\
  \bibinfo {pages} {042335} (\bibinfo {year} {2012})}\BibitemShut {NoStop}%
\bibitem [{\citenamefont {Wei}\ \emph {et~al.}(2016)\citenamefont {Wei},
  \citenamefont {Ruan},\ and\ \citenamefont {Long}}]{Wei-sr-2016}%
  \BibitemOpen
  \bibfield  {author} {\bibinfo {author} {\bibfnamefont {S.-J.}\ \bibnamefont
  {Wei}}, \bibinfo {author} {\bibfnamefont {D.}~\bibnamefont {Ruan}}, \ and\
  \bibinfo {author} {\bibfnamefont {G.-L.}\ \bibnamefont {Long}},\ }\href
  {\doibase 10.1038/srep30727} {\bibfield  {journal} {\bibinfo  {journal} {Sci.
  Rep.}\ }\textbf {\bibinfo {volume} {6}},\ \bibinfo {pages} {30727} (\bibinfo
  {year} {2016})}\BibitemShut {NoStop}%
\bibitem [{\citenamefont {Zheng}(2021)}]{Zheng-sr-2021}%
  \BibitemOpen
  \bibfield  {author} {\bibinfo {author} {\bibfnamefont {C.}~\bibnamefont
  {Zheng}},\ }\href {\doibase 10.1038/s41598-021-83521-5} {\bibfield  {journal}
  {\bibinfo  {journal} {Sci. Rep.}\ }\textbf {\bibinfo {volume} {11}},\
  \bibinfo {pages} {3960} (\bibinfo {year} {2021})}\BibitemShut {NoStop}%
\bibitem [{\citenamefont {Di~Candia}\ \emph {et~al.}(2015)\citenamefont
  {Di~Candia}, \citenamefont {Pedernales}, \citenamefont {del Campo},
  \citenamefont {Solano},\ and\ \citenamefont {Casanova}}]{Candia-sr-2015}%
  \BibitemOpen
  \bibfield  {author} {\bibinfo {author} {\bibfnamefont {R.}~\bibnamefont
  {Di~Candia}}, \bibinfo {author} {\bibfnamefont {J.~S.}\ \bibnamefont
  {Pedernales}}, \bibinfo {author} {\bibfnamefont {A.}~\bibnamefont {del
  Campo}}, \bibinfo {author} {\bibfnamefont {E.}~\bibnamefont {Solano}}, \ and\
  \bibinfo {author} {\bibfnamefont {J.}~\bibnamefont {Casanova}},\ }\href
  {\doibase 10.1038/srep09981} {\bibfield  {journal} {\bibinfo  {journal} {Sci.
  Rep.}\ }\textbf {\bibinfo {volume} {5}},\ \bibinfo {pages} {9981} (\bibinfo
  {year} {2015})}\BibitemShut {NoStop}%
\bibitem [{\citenamefont {Zhang}\ \emph {et~al.}(2021)\citenamefont {Zhang},
  \citenamefont {Tao}, \citenamefont {He}, \citenamefont {Chen}, \citenamefont
  {Kong}, \citenamefont {Deng}, \citenamefont {Lambert},\ and\ \citenamefont
  {Ai}}]{Zhang-fp-2021}%
  \BibitemOpen
  \bibfield  {author} {\bibinfo {author} {\bibfnamefont {N.-N.}\ \bibnamefont
  {Zhang}}, \bibinfo {author} {\bibfnamefont {M.-J.}\ \bibnamefont {Tao}},
  \bibinfo {author} {\bibfnamefont {W.-T.}\ \bibnamefont {He}}, \bibinfo
  {author} {\bibfnamefont {X.-Y.}\ \bibnamefont {Chen}}, \bibinfo {author}
  {\bibfnamefont {X.-Y.}\ \bibnamefont {Kong}}, \bibinfo {author}
  {\bibfnamefont {F.-G.}\ \bibnamefont {Deng}}, \bibinfo {author}
  {\bibfnamefont {N.}~\bibnamefont {Lambert}}, \ and\ \bibinfo {author}
  {\bibfnamefont {Q.}~\bibnamefont {Ai}},\ }\href {\doibase
  10.1007/s11467-021-1064-y} {\bibfield  {journal} {\bibinfo  {journal} {Front.
  Phys.}\ }\textbf {\bibinfo {volume} {16}},\ \bibinfo {pages} {51501}
  (\bibinfo {year} {2021})}\BibitemShut {NoStop}%
\bibitem [{\citenamefont {Liu}\ \emph {et~al.}(2011)\citenamefont {Liu},
  \citenamefont {Li}, \citenamefont {Huang}, \citenamefont {Li}, \citenamefont
  {Guo}, \citenamefont {Laine}, \citenamefont {Breuer},\ and\ \citenamefont
  {Piilo}}]{Liu-np-2011}%
  \BibitemOpen
  \bibfield  {author} {\bibinfo {author} {\bibfnamefont {B.-H.}\ \bibnamefont
  {Liu}}, \bibinfo {author} {\bibfnamefont {L.}~\bibnamefont {Li}}, \bibinfo
  {author} {\bibfnamefont {Y.-F.}\ \bibnamefont {Huang}}, \bibinfo {author}
  {\bibfnamefont {C.-F.}\ \bibnamefont {Li}}, \bibinfo {author} {\bibfnamefont
  {G.-C.}\ \bibnamefont {Guo}}, \bibinfo {author} {\bibfnamefont {E.-M.}\
  \bibnamefont {Laine}}, \bibinfo {author} {\bibfnamefont {H.-P.}\ \bibnamefont
  {Breuer}}, \ and\ \bibinfo {author} {\bibfnamefont {J.}~\bibnamefont
  {Piilo}},\ }\href {\doibase 10.1038/nphys2085} {\bibfield  {journal}
  {\bibinfo  {journal} {Nat. Phys.}\ }\textbf {\bibinfo {volume} {7}},\
  \bibinfo {pages} {931} (\bibinfo {year} {2011})}\BibitemShut {NoStop}%
\bibitem [{\citenamefont {Bernardes}\ \emph {et~al.}(2016)\citenamefont
  {Bernardes}, \citenamefont {Peterson}, \citenamefont {Sarthour},
  \citenamefont {Souza}, \citenamefont {Monken}, \citenamefont {Roditi},
  \citenamefont {Oliveira},\ and\ \citenamefont {Santos}}]{Bernardes-sr-2016}%
  \BibitemOpen
  \bibfield  {author} {\bibinfo {author} {\bibfnamefont {N.~K.}\ \bibnamefont
  {Bernardes}}, \bibinfo {author} {\bibfnamefont {J.~P.~S.}\ \bibnamefont
  {Peterson}}, \bibinfo {author} {\bibfnamefont {R.~S.}\ \bibnamefont
  {Sarthour}}, \bibinfo {author} {\bibfnamefont {A.~M.}\ \bibnamefont {Souza}},
  \bibinfo {author} {\bibfnamefont {C.~H.}\ \bibnamefont {Monken}}, \bibinfo
  {author} {\bibfnamefont {I.}~\bibnamefont {Roditi}}, \bibinfo {author}
  {\bibfnamefont {I.~S.}\ \bibnamefont {Oliveira}}, \ and\ \bibinfo {author}
  {\bibfnamefont {M.~F.}\ \bibnamefont {Santos}},\ }\href {\doibase
  10.1038/srep33945} {\bibfield  {journal} {\bibinfo  {journal} {Sci. Rep.}\
  }\textbf {\bibinfo {volume} {6}},\ \bibinfo {pages} {33945} (\bibinfo {year}
  {2016})}\BibitemShut {NoStop}%
\bibitem [{\citenamefont {Liu}\ \emph {et~al.}(2018)\citenamefont {Liu},
  \citenamefont {Lyyra}, \citenamefont {Sun}, \citenamefont {Liu},
  \citenamefont {Li}, \citenamefont {Guo}, \citenamefont {Maniscalco},\ and\
  \citenamefont {Piilo}}]{Liu-natcom-2018}%
  \BibitemOpen
  \bibfield  {author} {\bibinfo {author} {\bibfnamefont {Z.-D.}\ \bibnamefont
  {Liu}}, \bibinfo {author} {\bibfnamefont {H.}~\bibnamefont {Lyyra}}, \bibinfo
  {author} {\bibfnamefont {Y.-N.}\ \bibnamefont {Sun}}, \bibinfo {author}
  {\bibfnamefont {B.-H.}\ \bibnamefont {Liu}}, \bibinfo {author} {\bibfnamefont
  {C.-F.}\ \bibnamefont {Li}}, \bibinfo {author} {\bibfnamefont {G.-C.}\
  \bibnamefont {Guo}}, \bibinfo {author} {\bibfnamefont {S.}~\bibnamefont
  {Maniscalco}}, \ and\ \bibinfo {author} {\bibfnamefont {J.}~\bibnamefont
  {Piilo}},\ }\href {\doibase 10.1038/s41467-018-05817-x} {\bibfield  {journal}
  {\bibinfo  {journal} {Nat. Commun.}\ }\textbf {\bibinfo {volume} {9}},\
  \bibinfo {pages} {3453} (\bibinfo {year} {2018})}\BibitemShut {NoStop}%
\bibitem [{\citenamefont {Patsch}\ \emph {et~al.}(2020)\citenamefont {Patsch},
  \citenamefont {Maniscalco},\ and\ \citenamefont {Koch}}]{patsch-prr-2020}%
  \BibitemOpen
  \bibfield  {author} {\bibinfo {author} {\bibfnamefont {S.}~\bibnamefont
  {Patsch}}, \bibinfo {author} {\bibfnamefont {S.}~\bibnamefont {Maniscalco}},
  \ and\ \bibinfo {author} {\bibfnamefont {C.~P.}\ \bibnamefont {Koch}},\
  }\href {\doibase 10.1103/PhysRevResearch.2.023133} {\bibfield  {journal}
  {\bibinfo  {journal} {Phys. Rev. Research}\ }\textbf {\bibinfo {volume}
  {2}},\ \bibinfo {pages} {023133} (\bibinfo {year} {2020})}\BibitemShut
  {NoStop}%
\bibitem [{\citenamefont {Garc{\'i}a-P{\'e}rez}\ \emph
  {et~al.}(2020)\citenamefont {Garc{\'i}a-P{\'e}rez}, \citenamefont {Rossi},\
  and\ \citenamefont {Maniscalco}}]{garcia-npj-2020}%
  \BibitemOpen
  \bibfield  {author} {\bibinfo {author} {\bibfnamefont {G.}~\bibnamefont
  {Garc{\'i}a-P{\'e}rez}}, \bibinfo {author} {\bibfnamefont {M.~A.~C.}\
  \bibnamefont {Rossi}}, \ and\ \bibinfo {author} {\bibfnamefont
  {S.}~\bibnamefont {Maniscalco}},\ }\href {\doibase 10.1038/s41534-019-0235-y}
  {\bibfield  {journal} {\bibinfo  {journal} {npj Quantum Inf.}\ }\textbf
  {\bibinfo {volume} {6}},\ \bibinfo {pages} {1} (\bibinfo {year}
  {2020})}\BibitemShut {NoStop}%
\bibitem [{\citenamefont {Dogra}\ \emph {et~al.}(2021)\citenamefont {Dogra},
  \citenamefont {Melnikov},\ and\ \citenamefont
  {Paraoanu}}]{dogra-commphy-2021}%
  \BibitemOpen
  \bibfield  {author} {\bibinfo {author} {\bibfnamefont {S.}~\bibnamefont
  {Dogra}}, \bibinfo {author} {\bibfnamefont {A.~A.}\ \bibnamefont {Melnikov}},
  \ and\ \bibinfo {author} {\bibfnamefont {G.~S.}\ \bibnamefont {Paraoanu}},\
  }\href {\doibase 10.1038/s42005-021-00534-2} {\bibfield  {journal} {\bibinfo
  {journal} {Commun. Phys.}\ }\textbf {\bibinfo {volume} {4}},\ \bibinfo
  {pages} {26} (\bibinfo {year} {2021})}\BibitemShut {NoStop}%
\bibitem [{\citenamefont {Shirokov}(2020)}]{shirokov-jmp-2020}%
  \BibitemOpen
  \bibfield  {author} {\bibinfo {author} {\bibfnamefont {M.~E.}\ \bibnamefont
  {Shirokov}},\ }\href {\doibase 10.1063/1.5134660} {\bibfield  {journal}
  {\bibinfo  {journal} {J. Math. Phys.}\ }\textbf {\bibinfo {volume} {61}},\
  \bibinfo {pages} {082204} (\bibinfo {year} {2020})},\ \Eprint
  {http://arxiv.org/abs/https://doi.org/10.1063/1.5134660}
  {https://doi.org/10.1063/1.5134660} \BibitemShut {NoStop}%
\bibitem [{\citenamefont {Head-Marsden}\ \emph {et~al.}(2021)\citenamefont
  {Head-Marsden}, \citenamefont {Krastanov}, \citenamefont {Mazziotti},\ and\
  \citenamefont {Narang}}]{prineha-prr-2021}%
  \BibitemOpen
  \bibfield  {author} {\bibinfo {author} {\bibfnamefont {K.}~\bibnamefont
  {Head-Marsden}}, \bibinfo {author} {\bibfnamefont {S.}~\bibnamefont
  {Krastanov}}, \bibinfo {author} {\bibfnamefont {D.~A.}\ \bibnamefont
  {Mazziotti}}, \ and\ \bibinfo {author} {\bibfnamefont {P.}~\bibnamefont
  {Narang}},\ }\href {\doibase 10.1103/PhysRevResearch.3.013182} {\bibfield
  {journal} {\bibinfo  {journal} {Phys. Rev. Research}\ }\textbf {\bibinfo
  {volume} {3}},\ \bibinfo {pages} {013182} (\bibinfo {year}
  {2021})}\BibitemShut {NoStop}%
\bibitem [{\citenamefont {Hu}\ \emph {et~al.}(2020)\citenamefont {Hu},
  \citenamefont {Xia},\ and\ \citenamefont {Kais}}]{Hu-sr-2020}%
  \BibitemOpen
  \bibfield  {author} {\bibinfo {author} {\bibfnamefont {Z.}~\bibnamefont
  {Hu}}, \bibinfo {author} {\bibfnamefont {R.}~\bibnamefont {Xia}}, \ and\
  \bibinfo {author} {\bibfnamefont {S.}~\bibnamefont {Kais}},\ }\href {\doibase
  10.1038/s41598-020-60321-x} {\bibfield  {journal} {\bibinfo  {journal} {Sci.
  Rep.}\ }\textbf {\bibinfo {volume} {10}},\ \bibinfo {pages} {3301} (\bibinfo
  {year} {2020})}\BibitemShut {NoStop}%
\bibitem [{\citenamefont {Gaikwad}\ \emph {et~al.}(2018)\citenamefont
  {Gaikwad}, \citenamefont {Rehal}, \citenamefont {Singh}, \citenamefont
  {Arvind},\ and\ \citenamefont {Dorai}}]{gaikwad-pra-2018}%
  \BibitemOpen
  \bibfield  {author} {\bibinfo {author} {\bibfnamefont {A.}~\bibnamefont
  {Gaikwad}}, \bibinfo {author} {\bibfnamefont {D.}~\bibnamefont {Rehal}},
  \bibinfo {author} {\bibfnamefont {A.}~\bibnamefont {Singh}}, \bibinfo
  {author} {\bibnamefont {Arvind}}, \ and\ \bibinfo {author} {\bibfnamefont
  {K.}~\bibnamefont {Dorai}},\ }\href {\doibase 10.1103/PhysRevA.97.022311}
  {\bibfield  {journal} {\bibinfo  {journal} {Phys. Rev. A}\ }\textbf {\bibinfo
  {volume} {97}},\ \bibinfo {pages} {022311} (\bibinfo {year}
  {2018})}\BibitemShut {NoStop}%
\bibitem [{\citenamefont {Gaikwad}\ \emph
  {et~al.}(2021{\natexlab{a}})\citenamefont {Gaikwad}, \citenamefont {Shende},\
  and\ \citenamefont {Dorai}}]{gaikwad-ijqi-2020}%
  \BibitemOpen
  \bibfield  {author} {\bibinfo {author} {\bibfnamefont {A.}~\bibnamefont
  {Gaikwad}}, \bibinfo {author} {\bibfnamefont {K.}~\bibnamefont {Shende}}, \
  and\ \bibinfo {author} {\bibfnamefont {K.}~\bibnamefont {Dorai}},\ }\href
  {\doibase 10.1142/S0219749920400043} {\bibfield  {journal} {\bibinfo
  {journal} {International Journal of Quantum Information}\ }\textbf {\bibinfo
  {volume} {19}},\ \bibinfo {pages} {2040004} (\bibinfo {year}
  {2021}{\natexlab{a}})},\ \Eprint
  {http://arxiv.org/abs/https://doi.org/10.1142/S0219749920400043}
  {https://doi.org/10.1142/S0219749920400043} \BibitemShut {NoStop}%
\bibitem [{\citenamefont {Gaikwad}\ \emph
  {et~al.}(2021{\natexlab{b}})\citenamefont {Gaikwad}, \citenamefont {Arvind},\
  and\ \citenamefont {Dorai}}]{gaikwad-arxiv-2021}%
  \BibitemOpen
  \bibfield  {author} {\bibinfo {author} {\bibfnamefont {A.}~\bibnamefont
  {Gaikwad}}, \bibinfo {author} {\bibnamefont {Arvind}}, \ and\ \bibinfo
  {author} {\bibfnamefont {K.}~\bibnamefont {Dorai}},\ }\href@noop {} {\enquote
  {\bibinfo {title} {Efficient experimental characterization of quantum
  processes via compressed sensing on an NMR quantum processor},}\ } (\bibinfo
  {year} {2021}{\natexlab{b}}),\ \Eprint {http://arxiv.org/abs/2109.13189}
  {arXiv:2109.13189 [quant-ph]} \BibitemShut {NoStop}%
\bibitem [{\citenamefont {Kraus}\ \emph {et~al.}(1983)\citenamefont {Kraus},
  \citenamefont {Bohm}, \citenamefont {Dollard},\ and\ \citenamefont
  {Wootters}}]{kraus-book-1983}%
  \BibitemOpen
  \bibfield  {author} {\bibinfo {author} {\bibfnamefont {K.}~\bibnamefont
  {Kraus}}, \bibinfo {author} {\bibfnamefont {A.}~\bibnamefont {Bohm}},
  \bibinfo {author} {\bibfnamefont {J.}~\bibnamefont {Dollard}}, \ and\
  \bibinfo {author} {\bibfnamefont {W.}~\bibnamefont {Wootters}},\ }\href
  {\doibase 10.1007/3-540-12732-1} {\emph {\bibinfo {title} {States, Effects,
  and Operations: Fundamental Notions of Quantum Theory}}}\ (\bibinfo
  {publisher} {Springer-Verlag Berlin Heidelberg},\ \bibinfo {year}
  {1983})\BibitemShut {NoStop}%
\bibitem [{\citenamefont {Oliveira}\ \emph {et~al.}(2007)\citenamefont
  {Oliveira}, \citenamefont {Bonagamba}, \citenamefont {Sarthour},
  \citenamefont {Freitas},\ and\ \citenamefont {deAzevedo}}]{oliveira-book-07}%
  \BibitemOpen
  \bibfield  {author} {\bibinfo {author} {\bibfnamefont {I.~S.}\ \bibnamefont
  {Oliveira}}, \bibinfo {author} {\bibfnamefont {T.~J.}\ \bibnamefont
  {Bonagamba}}, \bibinfo {author} {\bibfnamefont {R.~S.}\ \bibnamefont
  {Sarthour}}, \bibinfo {author} {\bibfnamefont {J.~C.~C.}\ \bibnamefont
  {Freitas}}, \ and\ \bibinfo {author} {\bibfnamefont {E.~R.}\ \bibnamefont
  {deAzevedo}},\ }\href@noop {} {\emph {\bibinfo {title} {NMR Quantum
  Information Processing}}}\ (\bibinfo  {publisher} {Elsevier},\ \bibinfo
  {address} {Linacre House, Jordan Hill, Oxford OX2 8DP, UK},\ \bibinfo {year}
  {2007})\BibitemShut {NoStop}%
\bibitem [{\citenamefont {Gaikwad}\ \emph
  {et~al.}(2021{\natexlab{c}})\citenamefont {Gaikwad}, \citenamefont
  {{Arvind}},\ and\ \citenamefont {Dorai}}]{Gaikwad-qip-2021}%
  \BibitemOpen
  \bibfield  {author} {\bibinfo {author} {\bibfnamefont {A.}~\bibnamefont
  {Gaikwad}}, \bibinfo {author} {\bibnamefont {{Arvind}}}, \ and\ \bibinfo
  {author} {\bibfnamefont {K.}~\bibnamefont {Dorai}},\ }\href {\doibase
  10.1007/s11128-020-02930-z} {\bibfield  {journal} {\bibinfo  {journal}
  {Quant. Inf. Proc.}\ }\textbf {\bibinfo {volume} {20}},\ \bibinfo {pages}
  {19} (\bibinfo {year} {2021}{\natexlab{c}})}\BibitemShut {NoStop}%
\bibitem [{\citenamefont {Singh}\ \emph
  {et~al.}(2017{\natexlab{a}})\citenamefont {Singh}, \citenamefont {Arvind},\
  and\ \citenamefont {Dorai}}]{singh-epl-2017}%
  \BibitemOpen
  \bibfield  {author} {\bibinfo {author} {\bibfnamefont {H.}~\bibnamefont
  {Singh}}, \bibinfo {author} {\bibnamefont {Arvind}}, \ and\ \bibinfo {author}
  {\bibfnamefont {K.}~\bibnamefont {Dorai}},\ }\href {\doibase
  10.1209/0295-5075/118/50001} {\bibfield  {journal} {\bibinfo  {journal}
  {{EPL} (Europhysics Letters)}\ }\textbf {\bibinfo {volume} {118}},\ \bibinfo
  {pages} {50001} (\bibinfo {year} {2017}{\natexlab{a}})}\BibitemShut {NoStop}%
\bibitem [{\citenamefont {Singh}\ \emph
  {et~al.}(2017{\natexlab{b}})\citenamefont {Singh}, \citenamefont {Arvind},\
  and\ \citenamefont {Dorai}}]{singh-pra-2017}%
  \BibitemOpen
  \bibfield  {author} {\bibinfo {author} {\bibfnamefont {H.}~\bibnamefont
  {Singh}}, \bibinfo {author} {\bibnamefont {Arvind}}, \ and\ \bibinfo {author}
  {\bibfnamefont {K.}~\bibnamefont {Dorai}},\ }\href {\doibase
  10.1103/PhysRevA.95.052337} {\bibfield  {journal} {\bibinfo  {journal} {Phys.
  Rev. A}\ }\textbf {\bibinfo {volume} {95}},\ \bibinfo {pages} {052337}
  (\bibinfo {year} {2017}{\natexlab{b}})}\BibitemShut {NoStop}%
\bibitem [{\citenamefont {Singh}\ \emph {et~al.}(2018)\citenamefont {Singh},
  \citenamefont {Arvind},\ and\ \citenamefont {Dorai}}]{singh-pra-2018}%
  \BibitemOpen
  \bibfield  {author} {\bibinfo {author} {\bibfnamefont {H.}~\bibnamefont
  {Singh}}, \bibinfo {author} {\bibnamefont {Arvind}}, \ and\ \bibinfo {author}
  {\bibfnamefont {K.}~\bibnamefont {Dorai}},\ }\href {\doibase
  10.1103/PhysRevA.97.022302} {\bibfield  {journal} {\bibinfo  {journal} {Phys.
  Rev. A}\ }\textbf {\bibinfo {volume} {97}},\ \bibinfo {pages} {022302}
  (\bibinfo {year} {2018})}\BibitemShut {NoStop}%
\bibitem [{\citenamefont {Singh}\ \emph {et~al.}(2020)\citenamefont {Singh},
  \citenamefont {{Arvind}},\ and\ \citenamefont {Dorai}}]{singh-pramana-2020}%
  \BibitemOpen
  \bibfield  {author} {\bibinfo {author} {\bibfnamefont {H.}~\bibnamefont
  {Singh}}, \bibinfo {author} {\bibnamefont {{Arvind}}}, \ and\ \bibinfo
  {author} {\bibfnamefont {K.}~\bibnamefont {Dorai}},\ }\href {\doibase
  10.1007/s12043-020-02027-3} {\bibfield  {journal} {\bibinfo  {journal}
  {Pramana}\ }\textbf {\bibinfo {volume} {94}},\ \bibinfo {pages} {160}
  (\bibinfo {year} {2020})}\BibitemShut {NoStop}%
\bibitem [{\citenamefont {Childs}\ \emph {et~al.}(2001)\citenamefont {Childs},
  \citenamefont {Chuang},\ and\ \citenamefont {Leung}}]{childs-pra-2001}%
  \BibitemOpen
  \bibfield  {author} {\bibinfo {author} {\bibfnamefont {A.~M.}\ \bibnamefont
  {Childs}}, \bibinfo {author} {\bibfnamefont {I.~L.}\ \bibnamefont {Chuang}},
  \ and\ \bibinfo {author} {\bibfnamefont {D.~W.}\ \bibnamefont {Leung}},\
  }\href {\doibase 10.1103/PhysRevA.64.012314} {\bibfield  {journal} {\bibinfo
  {journal} {Phys. Rev. A}\ }\textbf {\bibinfo {volume} {64}},\ \bibinfo
  {pages} {012314} (\bibinfo {year} {2001})}\BibitemShut {NoStop}%
\bibitem [{\citenamefont {Iten}\ \emph {et~al.}(2016)\citenamefont {Iten},
  \citenamefont {Colbeck}, \citenamefont {Kukuljan}, \citenamefont {Home},\
  and\ \citenamefont {Christandl}}]{iten-pra-2016}%
  \BibitemOpen
  \bibfield  {author} {\bibinfo {author} {\bibfnamefont {R.}~\bibnamefont
  {Iten}}, \bibinfo {author} {\bibfnamefont {R.}~\bibnamefont {Colbeck}},
  \bibinfo {author} {\bibfnamefont {I.}~\bibnamefont {Kukuljan}}, \bibinfo
  {author} {\bibfnamefont {J.}~\bibnamefont {Home}}, \ and\ \bibinfo {author}
  {\bibfnamefont {M.}~\bibnamefont {Christandl}},\ }\href {\doibase
  10.1103/PhysRevA.93.032318} {\bibfield  {journal} {\bibinfo  {journal} {Phys.
  Rev. A}\ }\textbf {\bibinfo {volume} {93}},\ \bibinfo {pages} {032318}
  (\bibinfo {year} {2016})}\BibitemShut {NoStop}%
\bibitem [{\citenamefont {Iten}\ \emph {et~al.}(2021)\citenamefont {Iten},
  \citenamefont {Reardon-Smith}, \citenamefont {Malvetti}, \citenamefont
  {Mondada}, \citenamefont {Pauvert}, \citenamefont {Redmond}, \citenamefont
  {Kohli},\ and\ \citenamefont {Colbeck}}]{iten-arxiv-2021}%
  \BibitemOpen
  \bibfield  {author} {\bibinfo {author} {\bibfnamefont {R.}~\bibnamefont
  {Iten}}, \bibinfo {author} {\bibfnamefont {O.}~\bibnamefont {Reardon-Smith}},
  \bibinfo {author} {\bibfnamefont {E.}~\bibnamefont {Malvetti}}, \bibinfo
  {author} {\bibfnamefont {L.}~\bibnamefont {Mondada}}, \bibinfo {author}
  {\bibfnamefont {G.}~\bibnamefont {Pauvert}}, \bibinfo {author} {\bibfnamefont
  {E.}~\bibnamefont {Redmond}}, \bibinfo {author} {\bibfnamefont {R.~S.}\
  \bibnamefont {Kohli}}, \ and\ \bibinfo {author} {\bibfnamefont
  {R.}~\bibnamefont {Colbeck}},\ }\href@noop {} {\enquote {\bibinfo {title}
  {Introduction to universalqcompiler},}\ } (\bibinfo {year} {2021}),\ \Eprint
  {http://arxiv.org/abs/1904.01072} {arXiv:1904.01072 [quant-ph]} \BibitemShut
  {NoStop}%
\bibitem [{\citenamefont {{Le Bihan}}\ and\ \citenamefont
  {Johansen-Berg}(2012)}]{denis-neuro-2012}%
  \BibitemOpen
  \bibfield  {author} {\bibinfo {author} {\bibfnamefont {D.}~\bibnamefont {{Le
  Bihan}}}\ and\ \bibinfo {author} {\bibfnamefont {H.}~\bibnamefont
  {Johansen-Berg}},\ }\href {\doibase
  https://doi.org/10.1016/j.neuroimage.2011.11.006} {\bibfield  {journal}
  {\bibinfo  {journal} {NeuroImage}\ }\textbf {\bibinfo {volume} {61}},\
  \bibinfo {pages} {324} (\bibinfo {year} {2012})}\BibitemShut {NoStop}%
\bibitem [{\citenamefont {Pag{\'e}s}\ \emph {et~al.}(2017)\citenamefont
  {Pag{\'e}s}, \citenamefont {Gilard}, \citenamefont {Martino},\ and\
  \citenamefont {Malet-Martino}}]{paga-analyst-2017}%
  \BibitemOpen
  \bibfield  {author} {\bibinfo {author} {\bibfnamefont {G.}~\bibnamefont
  {Pag{\'e}s}}, \bibinfo {author} {\bibfnamefont {V.}~\bibnamefont {Gilard}},
  \bibinfo {author} {\bibfnamefont {R.}~\bibnamefont {Martino}}, \ and\
  \bibinfo {author} {\bibfnamefont {M.}~\bibnamefont {Malet-Martino}},\ }\href
  {\doibase 10.1039/C7AN01031A} {\bibfield  {journal} {\bibinfo  {journal}
  {Analyst}\ }\textbf {\bibinfo {volume} {142}},\ \bibinfo {pages} {3771}
  (\bibinfo {year} {2017})}\BibitemShut {NoStop}%
\bibitem [{\citenamefont {Han}\ \emph {et~al.}(2021)\citenamefont {Han},
  \citenamefont {Bazak}, \citenamefont {Chen}, \citenamefont {Graham},
  \citenamefont {Washton}, \citenamefont {Hu}, \citenamefont {Murugesan},\ and\
  \citenamefont {Mueller}}]{han-chemmat-2021}%
  \BibitemOpen
  \bibfield  {author} {\bibinfo {author} {\bibfnamefont {K.~S.}\ \bibnamefont
  {Han}}, \bibinfo {author} {\bibfnamefont {J.~D.}\ \bibnamefont {Bazak}},
  \bibinfo {author} {\bibfnamefont {Y.}~\bibnamefont {Chen}}, \bibinfo {author}
  {\bibfnamefont {T.~R.}\ \bibnamefont {Graham}}, \bibinfo {author}
  {\bibfnamefont {N.~M.}\ \bibnamefont {Washton}}, \bibinfo {author}
  {\bibfnamefont {J.~Z.}\ \bibnamefont {Hu}}, \bibinfo {author} {\bibfnamefont
  {V.}~\bibnamefont {Murugesan}}, \ and\ \bibinfo {author} {\bibfnamefont
  {K.~T.}\ \bibnamefont {Mueller}},\ }\href {\doibase
  10.1021/acs.chemmater.1c02891} {\bibfield  {journal} {\bibinfo  {journal}
  {Chem. Mater.}\ }\textbf {\bibinfo {volume} {33}},\ \bibinfo {pages} {8562}
  (\bibinfo {year} {2021})},\ \Eprint
  {http://arxiv.org/abs/https://doi.org/10.1021/acs.chemmater.1c02891}
  {https://doi.org/10.1021/acs.chemmater.1c02891} \BibitemShut {NoStop}%
\bibitem [{\citenamefont {Peterson}\ \emph {et~al.}(2020)\citenamefont
  {Peterson}, \citenamefont {Katiyar},\ and\ \citenamefont
  {Laflamme}}]{laflamme-arxiv-2020}%
  \BibitemOpen
  \bibfield  {author} {\bibinfo {author} {\bibfnamefont {J.~P.~S.}\
  \bibnamefont {Peterson}}, \bibinfo {author} {\bibfnamefont {H.}~\bibnamefont
  {Katiyar}}, \ and\ \bibinfo {author} {\bibfnamefont {R.}~\bibnamefont
  {Laflamme}},\ }\href@noop {} {\enquote {\bibinfo {title} {Fast simulation of
  magnetic field gradients for optimization of pulse sequences},}\ } (\bibinfo
  {year} {2020}),\ \Eprint {http://arxiv.org/abs/2006.10133} {arXiv:2006.10133
  [quant-ph]} \BibitemShut {NoStop}%
\bibitem [{\citenamefont {Devra}\ \emph {et~al.}(2018)\citenamefont {Devra},
  \citenamefont {Prabhu}, \citenamefont {Singh}, \citenamefont {{Arvind}},\
  and\ \citenamefont {Dorai}}]{Devra-qip-2018}%
  \BibitemOpen
  \bibfield  {author} {\bibinfo {author} {\bibfnamefont {A.}~\bibnamefont
  {Devra}}, \bibinfo {author} {\bibfnamefont {P.}~\bibnamefont {Prabhu}},
  \bibinfo {author} {\bibfnamefont {H.}~\bibnamefont {Singh}}, \bibinfo
  {author} {\bibnamefont {{Arvind}}}, \ and\ \bibinfo {author} {\bibfnamefont
  {K.}~\bibnamefont {Dorai}},\ }\href {\doibase 10.1007/s11128-018-1835-8}
  {\bibfield  {journal} {\bibinfo  {journal} {Quant. Inf. Proc.}\ }\textbf
  {\bibinfo {volume} {17}},\ \bibinfo {pages} {67} (\bibinfo {year}
  {2018})}\BibitemShut {NoStop}%
\end{thebibliography}

%
\begin{widetext}
\appendix
\section{Kraus operators \& unitary dilation operators for
phase damping channel}
\label{appendixA}
The complete set of Kraus operators corresponding to
an independent phase damping channel, acting on the two-qubit system
with parameter values $\gamma_1 = 1.4$, $\gamma_2 = 1.5$ and
$t=2$ sec, is given below:


\[  
 A_1= \left(
\begin{array}{cccc}
 -0.4723+0. i & 0.\, +0. i & 0.\, +0. i & 0.\, +0. i \\
 0.\, +0. i & 0.4723\, +0. i & 0.\, +0. i & 0.\, +0. i \\
 0.\, +0. i & 0.\, +0. i & 0.4723\, +0. i & 0.\, +0. i \\
 0.\, +0. i & 0.\, +0. i & 0.\, +0. i & -0.4723+0. i \\
\end{array}
\right)
\]

\[  
 A_2= \left(
\begin{array}{cccc}
 0.0181\, -0.4961 i & 0.\, +0. i & 0.\, +0. i & 0.\, +0. i \\
 0.\, +0. i & 0.0181\, -0.4961 i & 0.\, +0. i & 0.\, +0. i \\
 0.\, +0. i & 0.\, +0. i & -0.0181+0.4961 i & 0.\, +0. i \\
 0.\, +0. i & 0.\, +0. i & 0.\, +0. i & -0.0181+0.4961 i \\
\end{array}
\right)
\]

\[  
 A_3= \left(
\begin{array}{cccc}
 -0.0085-0.5019 i & 0.\, +0. i & 0.\, +0. i & 0.\, +0. i \\
 0.\, +0. i & 0.0085\, +0.5019 i & 0.\, +0. i & 0.\, +0. i \\
 0.\, +0. i & 0.\, +0. i & -0.0085-0.5019 i & 0.\, +0. i \\
 0.\, +0. i & 0.\, +0. i & 0.\, +0. i & 0.0085\, +0.5019 i \\
\end{array}
\right)
\]

\[  
 A_4= \left(
\begin{array}{cccc}
 -0.5276-0.007 i & 0.\, +0. i & 0.\, +0. i & 0.\, +0. i \\
 0.\, +0. i & -0.5276-0.007 i & 0.\, +0. i & 0.\, +0. i \\
 0.\, +0. i & 0.\, +0. i & -0.5276-0.007 i & 0.\, +0. i \\
 0.\, +0. i & 0.\, +0. i & 0.\, +0. i & -0.5276-0.007 i \\
\end{array}
\right)
\]

The decomposition of the unitary dilation operators $U_{A_i}$
corresponding to the Kraus operators  for the phase damping channel
are given below. We have used
the column-by-column decomposition method to decompose a given
unitary into single-qubit rotation gates and two-qubit CNOT gates.

\begin{enumerate}
\item  $U_{A_1}$: 
$^1R_{\bar{x}}^{\pi}$.$^1R_{\bar{y}}^{\frac{\pi}{2}}$.$U_{\rm{CNN}}$.${\rm
CNOT}_{32}$.$^2R_{\bar{z}}^{\frac{\pi}{2}}$.${\rm
CNOT}_{32}$.$^1R_{\bar{x}}^{\pi}$.$^1R_{\bar{y}}^{\theta_1}$.$U_{\rm{CNN}}$.$^1R_{\bar{x}}^{\pi}$.$^1R_{\bar{z}}^{\frac{3\pi}{2}}$.$U_{\rm{CNN}}$.$^1R_{\bar{z}}^{\frac{\pi}{2}}$
\\ where $U_{\rm{CNN}} = {\rm CNOT}_{31}.{\rm CNOT}_{21}$ 
and $\theta_1 = 0.5870$

\item  $U_{A_2}$ =
$^1R_{\bar{x}}^{\pi}$.$^1R_{\bar{y}}^{\frac{
\pi}{2}}$.${\rm CNOT}_{21}$.$^1R_{\bar{x}}^{\theta_3}
$.$^1R_{\bar{y}}^{\theta_2}$.
${\rm CNOT}_{21}$.$^1R_{\bar{x}}^{\theta_1}$.$^1R_{\bar{z}}^{\frac{
\pi}{2}}$.${\rm CNOT}_{21}$.$^2R_{\bar{z}}^{\frac{3 \pi}{2}}$
\\ where $\theta_1 = 3.0803 $, $\theta_2 = 0.5329 $, and
$\theta_3 = 1.6059 $

\item  $U_{A_3}$ =
$^1R_{\bar{x}}^{\pi}$.$^1R_{\bar{y}}^{\frac{
\pi}{2}}$.${\rm CNOT}_{31}$.$^1R_{\bar{x}}^{\theta_3}$.
$^1R_{\bar{y}}^{\theta_2}$.
${\rm CNOT}_{31}$.$^1R_{\bar{x}}^{\theta_1}$.$^1R_{\bar{z}}^{\frac{
\pi}{2}}$.${\rm CNOT}_{31}$.$^3R_{\bar{z}}^{\frac{3 \pi}{2}}$
\\ where $\theta_1 = 3.1711$, $\theta_2 = 0.5193 $, and
$\theta_3 = 1.5536 $

\item $U_{A_4}$ =
$^1R_{\bar{z}}^{\theta_3}$.$^1R_{\bar{y}}^{\theta_2}$.$^1R_{\bar{z}}^{\theta_1}$
where $\theta_1 = 3.1549$, $\theta_2 = 2.0299 $, and
$\theta_3 = 0.0133 $

\end{enumerate}
where $^iR_{\phi}^{\theta}$ represents a single-qubit rotation
gate acting on the $i$th qubit with the rotation angle $\theta$ and
the rotation axis is denoted by $\phi$ and ${\rm CNOT}_{ij}$
represents a two-qubit CNOT gate with the 
$i$th qubit being the control and the $j$th qubit 
being the target qubit.

\section{Kraus operators and unitary dilation operators for shaped MFGP}
\label{appendixB}
The complete set of Kraus operators corresponding to 
the desired shaped gradient
pulse with parameter values given in Section\ref{mfgp} applied on 
a two-qubit 
system were experimentally computed via 
the convex optimization based QPT method. The Kraus operators are 
given by:

\[
A_1=\left(
\begin{array}{cccc}
 0.1231\, -0.0877 i & -0.0038+0.0026 i & -0.0077+0.0085 i & 0.0023\, +0.0004 i \\
 0.0122\, -0.0279 i & -0.1899-0.1181 i & 0.0101\, +0.0085 i & 0.0097\, +0.006 i \\
 -0.0174+0.0165 i & -0.0073+0.0042 i & -0.3573+0.4876 i & 0.0167\, -0.0073 i \\
 -0.0036-0.0034 i & -0.0056+0.0133 i & -0.0009+0.0275 i & 0.5454\, +0.4572 i \\
\end{array}
\right)
\]

\[
A_2=\left(
\begin{array}{cccc}
 -0.0434-0.4568 i & 0.0061\, +0.0085 i & 0.0095\, +0.0121 i & -0.0055-0.0064 i \\
 0.0329\, +0.0096 i & 0.181\, -0.4594 i & -0.0029+0.0105 i & -0.0003+0.0002 i \\
 0.0017\, -0.0235 i & 0.0036\, -0.003 i & -0.35-0.3762 i & 0.0141\, -0.0184 i \\
 -0.0055-0.0042 i & 0.0124\, -0.007 i & 0.012\, +0.0275 i & 0.3231\, -0.3787 i \\
\end{array}
\right)
\]

\[
A_3=\left(
\begin{array}{cccc}
 -0.4842-0.5645 i & 0.0305\, +0.0057 i & 0.027\, -0.0027 i & -0.0011+0.0033 i \\
 -0.0206+0.0166 i & -0.327+0.0929 i & 0.0007\, -0.0019 i & 0.0034\, -0.0026 i \\
 0.0102\, +0.0216 i & -0.0024+0.0064 i & 0.3035\, -0.2407 i & 0.0096\, +0.0199 i \\
 -0.0005-0.0058 i & 0.0024\, +0.0041 i & 0.015\, +0.006 i & -0.0094+0.4166 i \\
\end{array}
\right)
\]

\[
A_4=\left(
\begin{array}{cccc}
 0.4475\, +0.0416 i & -0.0139+0.0256 i & -0.0099+0.0021 i & 0.0055\, +0.0044 i \\
 -0.0239-0.0035 i & -0.7081+0.2924 i & -0.0143-0.0018 i & 0.0063\, -0.0201 i \\
 0.0027\, -0.0084 i & 0.0055\, +0.0079 i & -0.1662-0.4034 i & 0.0107\, -0.0154 i \\
 0.0045\, -0.0062 i & 0.0167\, -0.0093 i & -0.0253+0.0106 i & 0.1022\, -0.1527 i \\
\end{array}
\right)
\]

The decomposition of unitary dilation operators $U_{A_i}$ corresponding to
respective Kraus operators are given below for a shaped gradient pulse.  We used
the column-by-column decomposition method to decompose a given unitary into
single-qubit rotations and CNOT gates. It turns out that in the case of a shaped
gradient pulse, the form of decomposition of unitary dilation operators
corresponding to all Kraus operators is the same. The general form of the
decomposition of unitary dilations is denoted by $U$ and given below.

\begin{equation}
\label{c1}
\begin{split}
U  =
{}^1R_{\bar{x}}^{\theta_{17}}.^1R_{\bar{y}}^{\theta_{16}}.\rm{CNOT}_{31}.^1R_{\bar{x}}^{\theta_{15}}.^1R_{\bar{y}}^{\theta_{14}}.\rm{CNOT}_{21}.^1R_{\bar{x}}^{\theta_{13}}.^1R_{\bar{y}}^{\theta_{12}}.\rm{CNOT}_{31}.^1R_{\bar{x}}^{\theta_{11}}.^1R_{\bar{y}}^{\theta_{10}}.\rm{CNOT}_{21}.^1R_{\bar{x}}^{\theta_9}.\\
^1R_{\bar{y}}^{\theta_8}.
\rm{CNOT}_{31}.^1R_{\bar{x}}^{\theta_7}.^1R_{\bar{y}}^{\theta_6}.\rm{CNOT}_{31}.^1R_{\bar{x}}^{\theta_5}.^1R_{\bar{z}}^{\theta_4}.\rm{CNOT}_{21}.^1R_{\bar{z}}^{\theta_3}.\rm{CNOT}_{31}.^1R_{\bar{z}}^{\theta_2}.\rm{CNOT}_{21}.^1R_{\bar{z}}^{\theta_1}.^3R_{\bar{z}}^{\theta_0}
\end{split}
\end{equation}

\begin{table}[h!] \caption{\label{table6} The values of $\theta_i$s
(Eq.\ref{c1}) required 
to the implement unitary dilation operators $U_{A_j}$.}
\begin{tabular}{c | c | c | c | c}
\hline \hline ~~~~~&~~~~~ $U_{A_1}$ ~~~~~& ~~~~~~$U_{A_2}$~~~~~~&~~~~~ $U_{A_3}$~~~~~~&~~~~~~$U_{A_4}$~~~~~\\
\hline \hline
$\theta_{0} $ &  1.5708   &  4.7124   &  4.7124  & 1.5708  \\
$\theta_{1} $ &  6.2759   &  0.0486   &  6.1354  & 0.1079  \\
$\theta_{2} $ &  5.7332   &  0.0306   &  0.1041  & 0.8518  \\
$\theta_{3} $ &  0.5359   &  0.1169   &  5.7425  & 5.6472  \\
$\theta_{4} $ &  4.2067   &  1.3599   &  2.4207  & 4.9160  \\
$\theta_{5} $ &  2.8192   &  3.0589   &  3.4918  & 2.6544  \\
$\theta_{6} $ &  1.8641   &  1.5181   &  1.2934  & 1.2327  \\
$\theta_{7} $ &  2.2842   &  1.0045   &  5.3556  & 1.0113  \\
$\theta_{8} $ &  0.4323   &  0.0979   &  0.4432  & 0.5851  \\
$\theta_{9} $ &  3.1416   &  3.1416   &  3.1416  & 3.1416  \\
$\theta_{10} $ &  0.4323   &  0.0979   &  0.4432  & 0.5851 \\
$\theta_{11} $ &  2.2856   &   0.6158  &  5.0076  & 3.7509  \\
$\theta_{12} $ &  1.0560   &  0.1859   &  0.6384  & 1.1481  \\
$\theta_{13} $ &  2.3100   &  2.9610   &  2.8007  & 5.0366  \\
$\theta_{14} $ &  0.6701   &  0.4389   &  0.9546  & 1.3207  \\
$\theta_{15} $ &  1.1972   &  0.1460   &  3.6109  & 4.3664  \\
$\theta_{16} $ &  1.6675   &  1.4041   &  2.5217  & 1.8259  \\
$\theta_{17} $ &  2.9373   &  2.1115   &  2.4411  & 3.8623  \\

\hline \end{tabular}
\end{table}
where $^iR_{\phi}^{\theta}$ represents a single-qubit rotation gate acting on
the $i$th qubit with rotation angle $\theta$ and axis of rotation 
$\phi$; $\rm{CNOT}_{ij}$ represents a standard two-qubit CNOT gate with 
$i$ being the control qubit 
and $j$ being the target qubit.

\end{widetext}
\end{document}